\documentclass[fleqn,usenatbib]{mnras}

\usepackage{newtxtext,bm}

\usepackage[T1]{fontenc}
\usepackage{url}

\DeclareRobustCommand{\VAN}[3]{#2}
\let\VANthebibliography\thebibliography
\def\thebibliography{\DeclareRobustCommand{\VAN}[3]{##3}\VANthebibliography}

\usepackage{graphicx}	
\usepackage{amsmath}	
\usepackage{amssymb}	

\title[The Formation history of the ONC]{SIRIUS Project. IV. The formation history of the Orion Nebula Cluster driven by clump mergers}

\author[M. S. Fujii et al.]{
Michiko S. Fujii,$^{1}$\thanks{E-mail: fujii@astron.s.u-tokyo.ac.jp (MSF)}
Long Wang,$^{2,1,3}$
Yutaka Hirai,$^{4,5}$\thanks{JSPS Research Fellow}
Yoshito Shimajiri,$^6$
Jun Kumamoto$^{1,7}$
and Takayuki Saitoh$^{8,9}$
\\
$^{1}$Department of Astronomy, Graduate School of Science, The University of Tokyo, 7-3-1 Hongo, Bunkyo-ku, Tokyo 113-0033, Japan\\
$^{2}$School of Physics and Astronomy, Sun Yat-sen University, Daxue Road, Zhuhai, 519082, China\\
$^{3}$RIKEN Center for Computational Science, 7-1-26 Minatojima-minami-machi, Chuo-ku,
Kobe, Hyogo 650-0047, Japan\\
$^{4}$Department of Physics and Astronomy, University of Notre Dame, 225 Nieuwland Science Hall, Notre Dame, 46556 IN, USA\\
$^{5}$Astronomical Institute, Tohoku University, 6-3 Aramaki, Aoba-ku, Sendai, Miyagi 980-8578, Japan\\
$^6$National Astronomical Observatory of Japan, 2-21-1 Osawa, Mitaka, Tokyo, 181-8588, Japan\\
$^7$Faculty of Business Administration, Chukyo Gakuin University, 1-104 Sendanbayashi, Nakatsugawa, Gifu 509-9195, Japan\\
$^8$Department of Planetology, Graduate School of Science, Kobe University, 1-1 Rokkodai-cho, Nada-ku, Kobe, Hyogo 657-8501, Japan\\
$^9$Earth-Life Science Institute, Tokyo Institute of Technology, 2-12-1 Ookayama, Meguro-ku, Tokyo 152-8551, Japan
}

\date{Accepted XXX. Received YYY; in original form ZZZ}

\pubyear{2022}

\begin{document}
\label{firstpage}
\pagerange{\pageref{firstpage}--\pageref{lastpage}}
\maketitle

\begin{abstract}
The Orion Nebula Cluster (ONC) is an excellent example for understanding the formation of star clusters. Recent studies have shown that ONC has three distinct age populations and anisotropy in velocity dispersions, which are key characteristics for understanding the formation history of the ONC. 
In this study, we perform a smoothed-particle hydrodynamics/$N$-body simulation of star cluster formation from a turbulent molecular cloud. In this simulation, stellar orbits are integrated using a high-order integrator without gravitational softening; therefore, we can follow the collisional evolution of star clusters. We find that hierarchical formation causes episodic star formation that is observed in the ONC. In our simulation, star clusters evolve due to mergers of subclumps. The mergers bring cold gas with the clumps into the forming cluster. This enhances the star formation in the cluster centre. The dense cold gas in the cluster centre continues to form stars until the latest time. This explains the compact distribution of the youngest stars observed in the ONC. Subclump mergers also contribute to the anisotropy in the velocity dispersions and the formation of runaway stars. However, the anisotropy disappears within 0.5\,Myr. The virial ratio of the cluster also increases after a merger due to the runaways. These results suggest that the ONC recently experienced a clump merger. We predict that most runaways originated from the ONC have already been found, but walkaways have not.
\end{abstract}




\section{Introduction}
The Orion Nebula Cluster (ONC) is one of the most massive young star clusters embedded in a molecular cloud in the Milky Way.  As the distance is relatively close to us the Earth ($\sim 390$\,pc) \citep{2017ApJ...834..142K,2019A&A...627A..57J}, detailed structures of the ONC have been investigated. Thus, the ONC is an excellent sample to understand the formation of massive open clusters.

The ONC has an estimated age of 1--3\,Myr \citep[e.g., ][]{1998ApJ...492..540H,2010ApJ...722.1092D}, and an age spread has been observed \citep[e.g., ][]{2010ApJ...722.1092D}. Recent studies have shown that there are three populations in the ONC, with the age difference of $\sim 1$\,Myr \citep{2017A&A...604A..22B,2019A&A...627A..57J,2019ApJ...870...32K}. One possible scenario explaining this age difference is the complete ejection of O-stars during the formation of the ONC. Once massive stars form in star clusters, their feedback halts star formation. However, massive stars in star clusters can be dynamically ejected from the cluster centre due to binary-single encounters, after which the cluster can start the star formation again \citep{2018A&A...612A..74K}. This scenario was tested with $N$-body simulations \citep{2019MNRAS.484.1843W}, which showed  moments without any O-type stars in the cluster centre. We note that, however, their simulations did not include the gas and star formation.

\citet{2014ApJ...787..109G} reported the age gradient from the ONC centre to the outer region. They observed that the mean stellar age of the central 0.15\,pc is $1.2\pm0.2$\,Myr, while that of the outer region is $1.9\pm 0.1$\,Myr. 
\citet{2014ApJ...787..109G} argued several possible cluster formation scenarios to generate the radial age gradient. One is that the most recent star formation occurred in the cluster centre. Stars can form in gas that is denser than a threshold density \citep{2010A&A...518L.102A}. If star clusters form in a globally collapsing cloud \citep{2000ApJ...540..255P}, star formation can continue in the cluster centre until the latest time because it is the densest. Another scenario is a hierarchical formation of star clusters \citep{2012ApJ...753...85F,2017MNRAS.467.1857B,2018NatAs...2..725H,2018MNRAS.481..688G,2020ApJ...891....2L,2021MNRAS.502.6157C}. If star clusters are formed via hierarchical mergers of subclumps or accretion of stars along with filaments, they can bring young stars into the cluster centre. Such hierarchical formation of clusters are often observed in numerical simulations of star cluster formation in turbulent molecular clouds \citep[e.g., ][]{2009MNRAS.392..590B,2017MNRAS.467.1313V,2018NatAs...2..725H,2018MNRAS.481..688G}.

Kinematic data are another important key to understanding the structure and formation history of the ONC. The first measurement of the velocity dispersion of the ONC was conducted  by \citet{1988AJ.....95.1755J}. 
\citet{2019AJ....157..109K} measured the velocity dispersions of the ONC as $(\sigma_{v_{\alpha}}, \sigma_{v_{\delta}})=(1.57\pm0.04, 2.12\pm0.06)$\,km\,s$^{-1}$ using Keck II and the Hubble Space Telescope (HST) assuming the distance of 400\,pc. \citet{2022ApJ...926..141T} reported the 3D velocity dispersions of 
$(\sigma_{v_{\alpha}}, \sigma_{v_{\delta}}, \sigma_{v_{r}}) = (1.64\pm0.12, 2.03\pm{0.13}, 2.56^{+0.16}_{-0.17})$\,km\,s$^{-1}$
also using Keck II and HST assuming the distance of 414\,pc \citep{2007A&A...474..515M}. 
Using Gaia data \citep{2018A&A...616A...1G}, \citet{2019ApJ...870...32K} measured the velocity dispersions of $(\sigma_{v_{\rm p1}}, \sigma_{v_{\rm p2}})=(2.2\pm0.2, 1.4\pm0.2)$\,km\,s$^{-1}$, where $\sigma_{v_{\rm p1}}$ and $\sigma_{v_{\rm p2}}$ are the velocity dispersions for the first and second principal axes, respectively.
These measurements show that the ONC is not spherical but has an elongated velocity structure. 

The virial ratio can be also calculated from the velocity dispersion measurements. \citet{2014ApJ...795...55D} estimated the velocity dispersion for the virial equilibrium to be $1.73$\,km\,s$^{-1}$. They estimated that the current virial ratio of the ONC is 0.9 using a velocity dispersion of $2.3$\,km\,s$^{-1}$, which obtained by \citet{1988AJ.....95.1755J}. As this measurement of the velocity dispersion is quite old, \citet{2019AJ....157..109K} estimated the virial ratio as $0.7\pm0.3$ using an updated velocity dispersion assuming the distance of 414\,pc, which is closer than  the assumption of 
 \citet{1988AJ.....95.1755J}.
These results suggest that the ONC is likely to be supervirial or virial.

\citet{2019AJ....157..109K} argued that the ONC is virialized and was formed in a turbulent molecular cloud through the several dynamical times of the cloud \citep{2006ApJ...641L.121T,2007ApJ...654..304K,2012ApJ...754...71K}. \citet{2014ApJ...795...55D} proposed that the ONC is in a superviral state as a result of recent gas dispersal.

Previous numerical studies have shown that star clusters are formed in turbulent molecular clouds. The simulations of star-cluster formation is becoming more realistic including various physics such as radiation, jets, stellar wind, and supernova \citep[e.g., STARFORGE project; ][]{2021MNRAS.506.3239G}.
We investigate the formation of the ONC through an $N$-body/hydrodynamics simulation. For the simulation, we adopted our newly developed $N$-body/smoothed-particle hydrodynamics (SPH) code, \textsc{asura+bridge} \citep{2021PASJ...73.1036H,2021PASJ...73.1057F,2021PASJ...73.1074F}. Unlike in previous studies, \textsc{asura+bridge} can integrate the stellar orbits without gravitational softening. This enables us to follow the dynamical evolution of star clusters as collisional systems, in which dynamical formation of binaries and close encounters play an important role. The ejections of massive stars caused by three-body encounters can also affect the ionization of gas. Using \textsc{ASURA+BRIDGE}, we aim to perform star-by-star simulations of star clusters resolving individual stellar orbits (SIRIUS project)\footnote{\url{https://sites.google.com/g.ecc.u-tokyo.ac.jp/sirius-project/}}.

In this study, we performed an $N$-body/SPH simulation for the formation of an ONC-like cluster. By comparing the results of the simulation with observations, we investigate the formation history of the ONC.

\section{Methods}
\subsection{Numerical simulation}

We performed an $N$-body/SPH simulation of star cluster formation using \textsc{asura+bridge} \citep{2021PASJ...73.1036H,2021PASJ...73.1057F,2021PASJ...73.1074F}, which is a program for star-by-star simulations based on an $N$-body/SPH
program \textsc{asura} \citep{2008PASJ...60..667S, 2009PASJ...61..481S}. 
With \textsc{asura+bridge}, we can integrate the motion of stars without softening lengths using high-order time integrators such as the sixth-order Hermite scheme \citep{2008NewA...13..498N} or the Particle-Particle Particle-Tree (P$^3$T) scheme \citep{2011PASJ...63..881O,2015ComAC...2....6I}. These high-order schemes are combined with our SPH code (tree code) using the Bridge scheme \citep{2007PASJ...59.1095F}, which is an extension of the mixed-variable symplectic scheme \citep{1991AJ....102.1528W,1991CeMDA..50...59K}.

For the P$^3$T scheme, \textsc{asura+bridge} adopts an $N$-body simulation code for star clusters, \textsc{petar} \citep{2020MNRAS.497..536W}. In \textsc{petar}, the forces from distant particles (soft force) is calculated using a tree \citep{Barnes1986} algorithm with a shared timestep (soft step) with a cutoff but the force from near-by particles (hard force) is integrated using a Hermite scheme with individual timesteps. \textsc{petar} is parallelized by using the  framework for developing particle simulators \citep[FDPS; ][]{Iwasawa2016FDPS}, which enables fast and scalable parallel calculations using supercomputers. \textsc{petar} includes the  slow-down algorithmic regularization (SDAR) scheme \citep{2020MNRAS.493.3398W}. With SDAR, \textsc{petar} can accurately integrate hard binaries, multiple systems, and close encounters.

In the Bridge scheme, star particles are kicked with the acceleration from gas potential every Bridge timestep ($\Delta t_{\rm B}$). This must be taken to be short enough to follow the change in the gas potential. In this simulation, we adopted $\Delta t_{\rm B}=200$\,yr following the results of test simulations carried out in \citet{2021PASJ...73.1057F}. 
In \textsc{petar}, we employ a timestep for the particle-tree (soft) part ($\Delta t_{\rm soft}$) to be $\Delta t_{\rm B}/1024$. These values are determined by following the results of \citet{2021PASJ...73.1057F} and \citet{2021PASJ...73.1074F}. We set the outer and inner cut-off radii for \textsc{petar} are $10^{-3}$ and $10^{-4}$\,pc, respectively.

\textsc{asura+bridge} adopts a probabilistic star formation \citep{2021PASJ...73.1036H}, in which star formation occurs following given star-formation probabilities depending on the local gas densities of individual gas particles and a given star-formation efficiency per local free-fall time ($c_{*}$). We applied this model for the formation of individual stars.
Once gas particles satisfy the following conditions for the star formation , namely,  (1) the density becomes higher than the threshold density ($n_{\rm{th}}$), (2) the temperature becomes  lower than the threshold temperature ($T_{\rm{th}}$), and (3) gas particles are converging ($\nabla\cdot\bm{v}<0$), a forming stellar mass is randomly drawn from the given mass function, the Kroupa mass function \citep {2001MNRAS.322..231K}. In this study, we utilize the chemical evolution library \citep[\textsc{CELib};][]{2017AJ....153...85S} to obtain eligible stellar masses. Then, the total gas mass within the maximum search radius ($r_{\rm{max}}$) is measured. If the drawn stellar mass exceeds half of the gas mass within $r_{\rm max}$, we again draw a stellar mass from the initial mass function (IMF). Once the stellar mass is determined, the newborn stellar particle is created from gas particles within $r_{\rm max}$. 
We reduce the mass of the all gas particles inside $r_{\rm max}$ to conserve the mass and create a new stellar particle. The position and velocity of the stellar particle is set to be the centre-of-mass position and velocity of the reduced gas. 
In our simulation, we set $r_{\rm max}=0.2$\,pc following \citet{2021PASJ...73.1036H}. We adopted $T_{\rm th}=20$\,K, $n_{\rm th}=7.4 \times 10^4$\,cm$^{-3}$, and $c_{*}=0.02$. For the mass function we adopted the upper- and lower-mass cut off of 0.1 and 150\,$M_{\odot}$, respectively. 

We used \textsc{Cloudy} ver.13.05 \citep{1998PASP..110..761F, 2013RMxAA..49..137F, 2017RMxAA..53..385F} for the radiative cooling and heating model for the interstellar medium and set the minimum temperature of the gas to be 20\,K. We assumed the solar metallicity is assumed for the model. 

We also adopted feedback from massive stars. We briefly summarize our scheme \citep[see][for more details]{2021PASJ...73.1074F}. Rather than solving radiation transfer, we modeled the H{\sc ii} region model around massive stars based on the Str{\"o}mgren radius \citep{1939ApJ....89..526S}. With ionizing photon rates emitted from massive stars \citep{2003ApJS..146..417L} and the local gas density, the Str{\"o}mgren radius for each massive star is calculated. Then, we add thermal energy to the gas particle inside the Str{\"o}mgren radius to increase their temperature to $10^4$\,K. We also provide mechanical feedback (outward velocity) to the gas particles within the Str{\"o}mgren radius. The gas velocity was determined following the model of \citet{2016ApJ...819..137K} for the radiation and that of \citet{2013MNRAS.436.1836R} for the stellar wind, but we excluded the term for the supernovae \citep{2013MNRAS.432..455D}. We ignored the mass loss due to the stellar wind, and therefore the stellar mass does not change during the simulation. We assumed the feedback for stars more massive than $10\,M_{\odot}$.

\subsection{Initial condition}
We performed a  series of $N$-body/SPH simulations of star-cluster formation starting from turbulent molecular clouds to pick up an eligible initial condition for our purpose. 
We adopted a homogeneous sphere of gas with a turbulent velocity field of which power spectrum has $\propto v^{-4}$ similar to that of \citet{2003MNRAS.343..413B}. After we tried some runs changing the initial mass, density, and virial ratio, we finally found an initial condition with the parameters summarized in Table \ref{tb:IC}, resulting in the formation of a star cluster similar to the ONC. We adopted the total gas mass of $M_{\rm g}=2\times 10^4M_{\odot}$, initial radius of $r_{g}=12$\, pc, and initial virial ratio, $\alpha_{\rm vir}=|E_{\rm k}|/|E_{\rm p}| = 0.45$, where $E_{\rm k}$ and $E_{\rm p}$ are kinetic and potential energy, respectively. 
The initial free-fall time of this model is 4.87\,Myr. 

We set the initial temperature of gas to 20\,K. We adopted a gas-particle mass ($m_{\rm g}$) of $0.01\,M_{\odot}$ and gas softening of $\epsilon_{\rm g}= 14,000$\,au (0.07\,pc). The initial distribution of the gas particles was constructed using the Astrophysical Multi-purpose Software Environment \citep[\textsc{amuse},][]{2013CoPhC.183..456P, 2013A&A...557A..84P, AMUSE}. 

\begin{table*}
\begin{center}
\caption{Model and parameters for the formation of the star cluster\label{tb:IC}}
\begin{tabular}{lccccccccccccc}
\hline
   Name  & $M_{\rm g}$  & $m_{\rm g}$ & $R_{\rm g}$& $n_{\rm ini}$ & $t_{\rm ff,ini}$  & $\alpha_{\rm vir}$ & $\epsilon_{\rm g}$ & $\epsilon_{\rm s}$ & $n_{\rm th}$ & $r_{\rm max}$ & $\Delta t_{\rm B}$ & $\Delta t_{\rm soft}$ & $r_{\rm out}$ \\
       & $(M_{\odot})$ & $(M_{\odot})$ & (pc) &  (cm$^{-3}$) & (Myr) &  & (pc) & (pc) & (cm$^{-3}$) & (pc) & (yr) & & (pc)\\ 
      \hline
  ONC & $2\times 10^4$ & $0.01$ & 12 & $79.9$ & $4.87$ & $0.45$  & $0.07$ & $0.0$ & $7.4\times 10^4$ & $0.2$ & $200$ & $\Delta t_{\rm B}/1024$ & 0.001 \\
\hline
\end{tabular}

From the left: model name, initial cloud mass ($M_{\rm g}$), gas-particle mass ($m_{\rm g}$), initial cloud radius ($R_{\rm g}$), initial cloud density ($n_{\rm ini}$), initial free-fall time ($t_{\rm ff, ini}$), initial virial ratio ($\alpha_{\rm vir}=|E_{\rm k}|/|E_{\rm p}|$), softening length for gas ($\epsilon_{\rm g}$) and stars ($\epsilon_{\rm s}$), star formation threshold density ($n_{\rm th}$), the maximum search radius for the star formation ($r_{\rm max}$), timestep for Bridge ($\Delta t_{\rm B}$) and soft part ($\Delta t_{\rm soft}$), and the outer cut-off radius for Bridge ($r_{\rm out}$).
\end{center}
\end{table*}

\section{Results}

\begin{table}
\begin{center}
\caption{Definition of three populations}
\begin{tabular}{lcc}
\hline
   Name  & Beccari {\it et. al.} (2017)  & This study\\
      \hline
  Very Young & 1.08--1.53\,Myr & 0.5--1.5\,Myr \\
 Young & 1.55--2.29\,Myr & 1.5--2.5\,Myr \\
 Old & 2.51--3.28\,Myr & 2.5--3.5\,Myr \\
      \hline
    \end{tabular}\label{tb:Age}
\end{center}
\end{table}

\begin{figure*}
    \includegraphics[width=7cm]{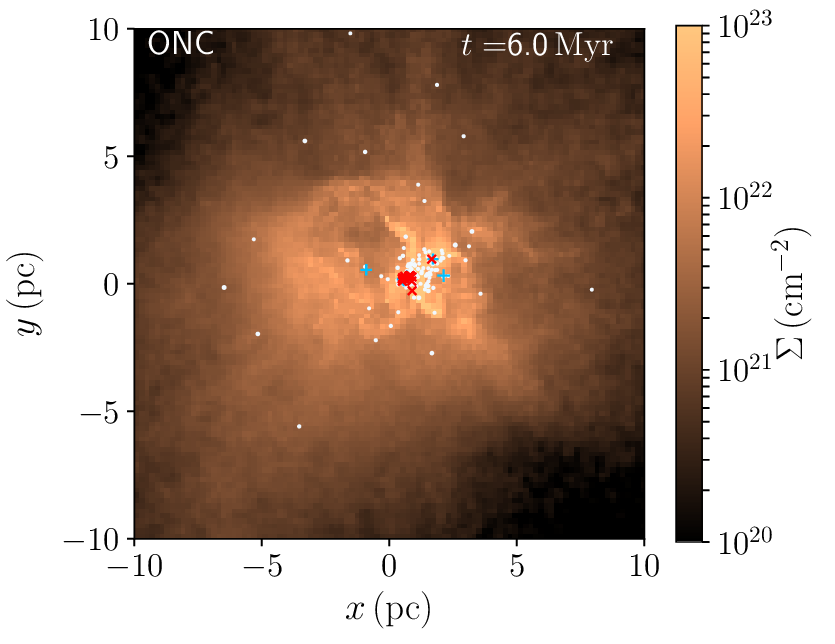}
	\includegraphics[width=7cm]{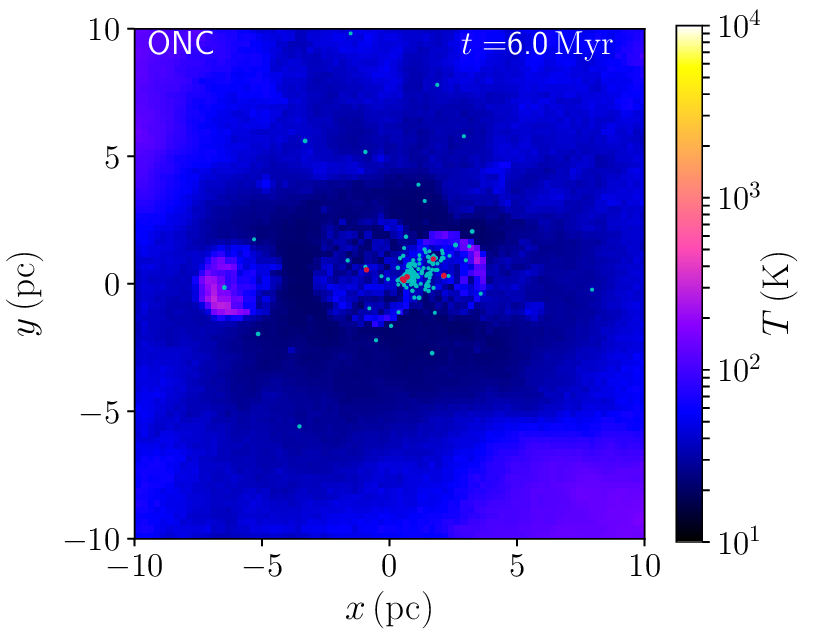}\\
    \includegraphics[width=7cm]{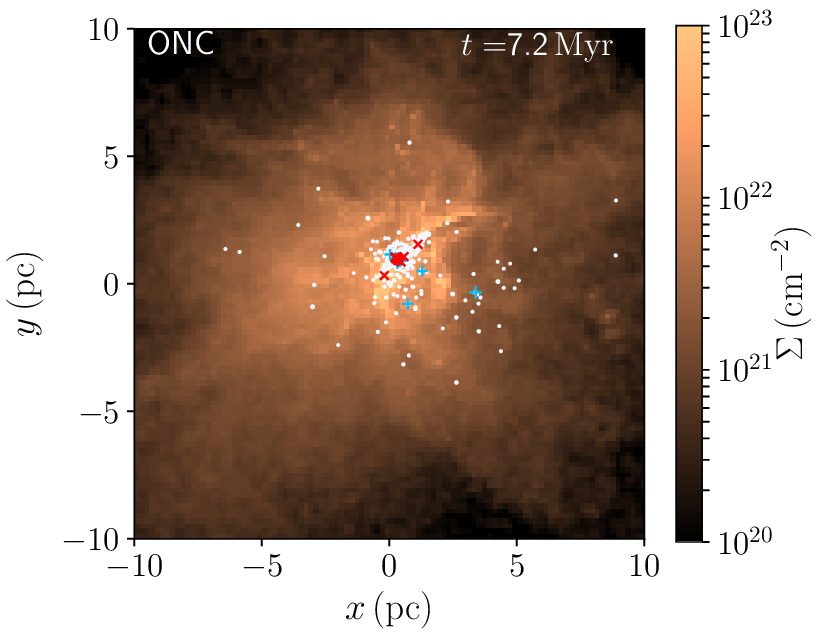}
	\includegraphics[width=7cm]{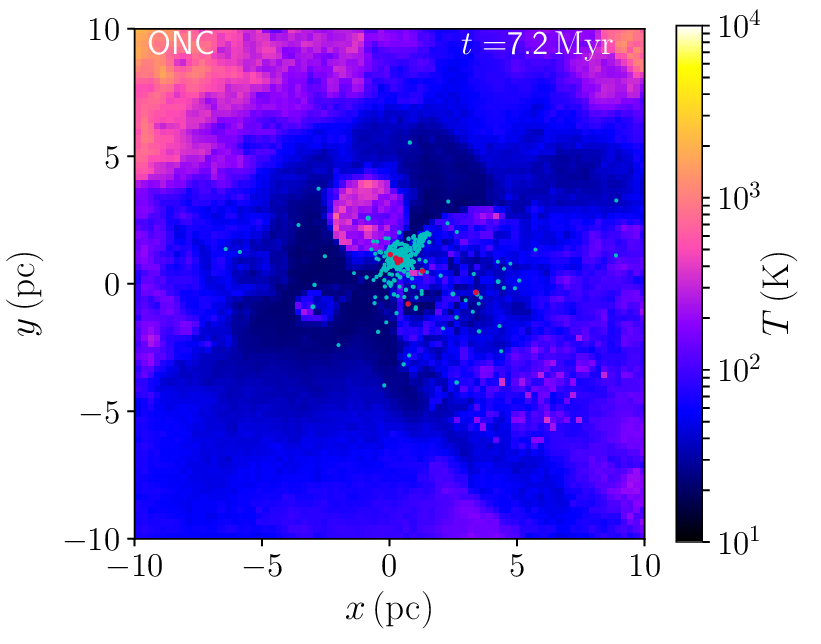}\\
	\includegraphics[width=7cm]{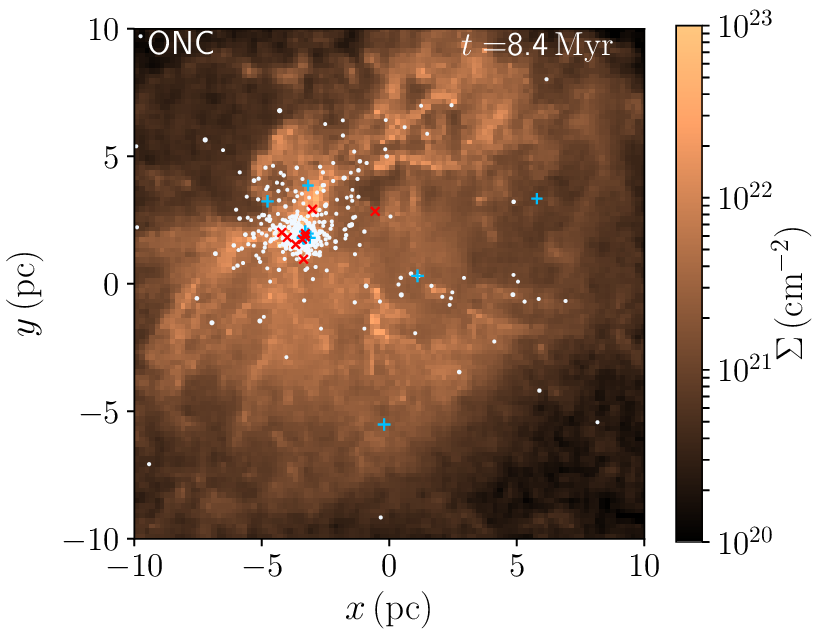}
	\includegraphics[width=7cm]{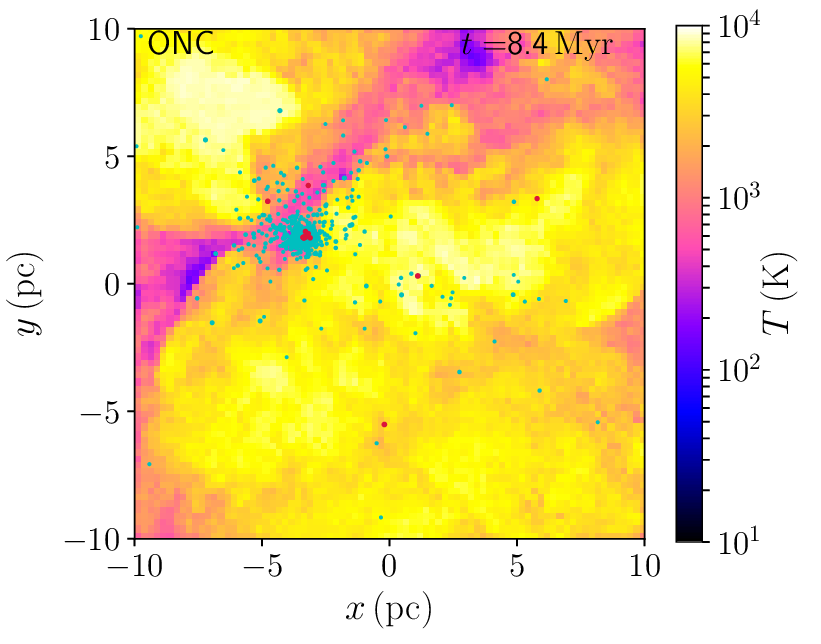}\\
	\includegraphics[width=7cm]{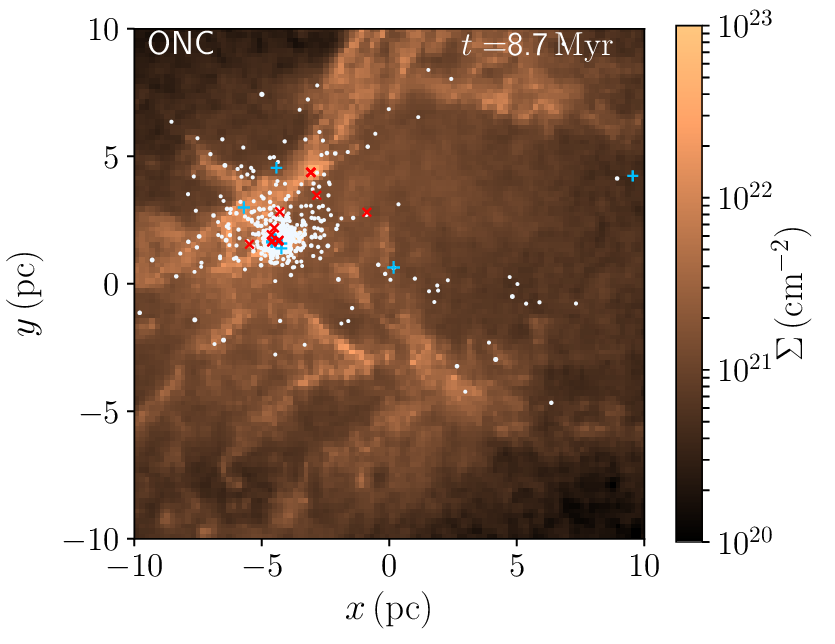}
	\includegraphics[width=7cm]{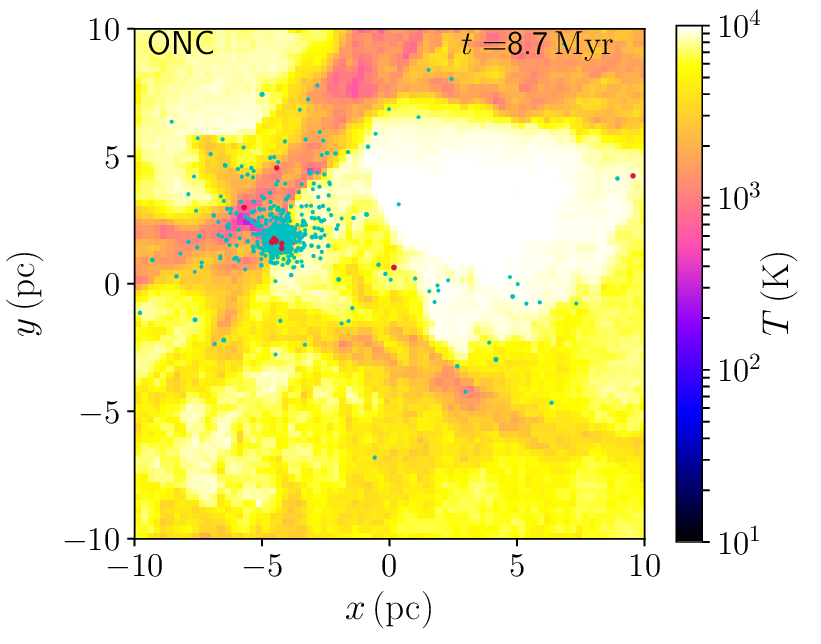}
    \caption{Snapshots of our simulation. Left: gas surface density and stars with $>1M_{\odot}$. Cyan pluses indicate massive stars with $>20M_{\odot}$ and red crossed indicate binaries. Right: gas temperature distribution and stars with $>1M_{\odot}$ (cyan dots). Red dots indicate massive stars. }
    \label{fig:snapshots}
\end{figure*}

\subsection{Cluster formation}
In Fig.~\ref{fig:snapshots}, we present the gas density and gas temperature distribution with stars. The initially homogeneous cloud collapses, forming filamentary structures because of the turbulence velocity field. The star formation starts in the dense regions at around one initial free-fall time ($t_{\rm ff}=4.87$ Myr in our model). 

\begin{figure}
	\includegraphics[width=0.95\columnwidth]{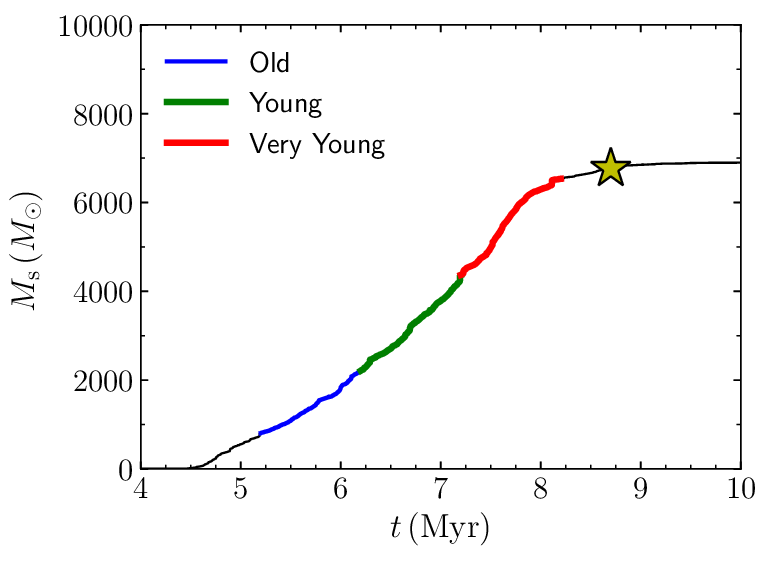}
	\caption{Time evolution of the total stellar mass. Colors indicate the periods that we defined as "Old" , "Young", "Very Young". Star indicate the "current time" (8.7\,Myr). }
    \label{fig:total_stellar_mass}
\end{figure}

The time evolution of the total stellar mass formed in the simulation is shown in Fig.~\ref{fig:total_stellar_mass}. The star formation accelerates until the feedback dominates at $\sim 8$\,Myr. Once the feedback starts to ionize the gas, star formation is quenched, and the residual gas is blown away from the cluster (see Fig.~\ref{fig:snapshots}). 

\begin{figure*}
    \includegraphics[width=7cm]{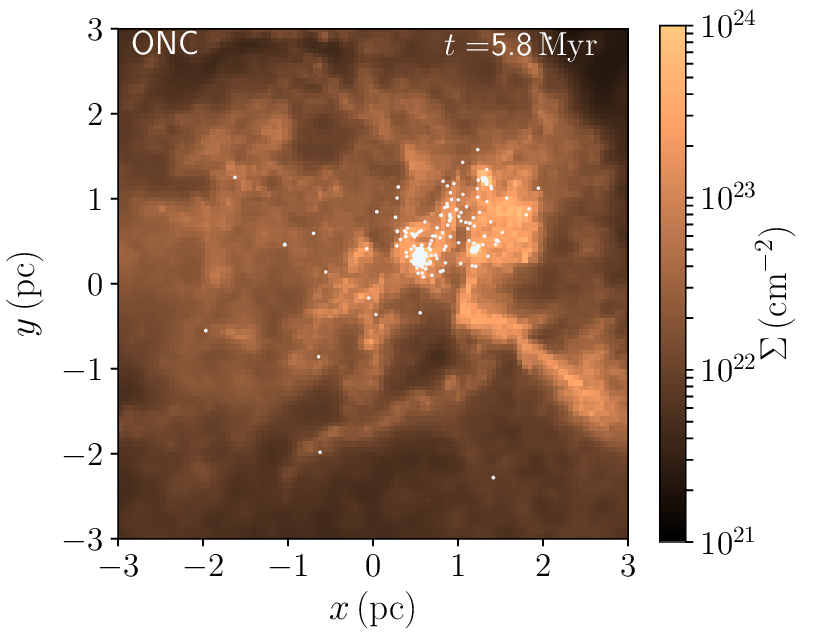}
    \includegraphics[width=7cm]{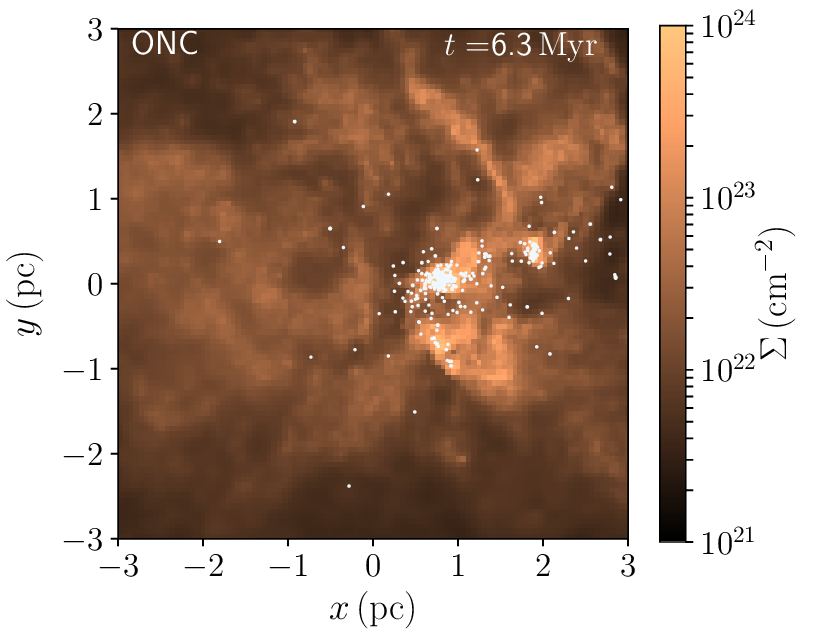}\\
    \includegraphics[width=7cm]{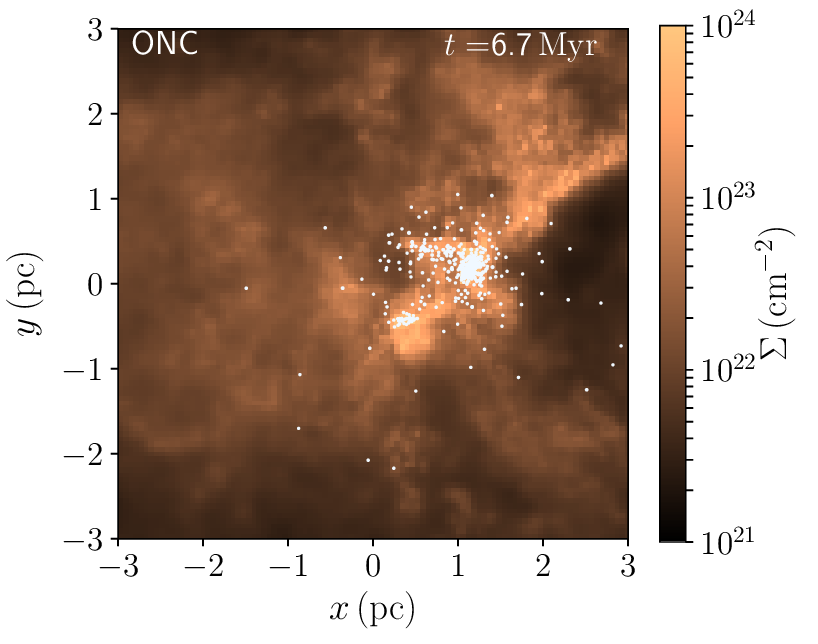}
	\includegraphics[width=7cm]{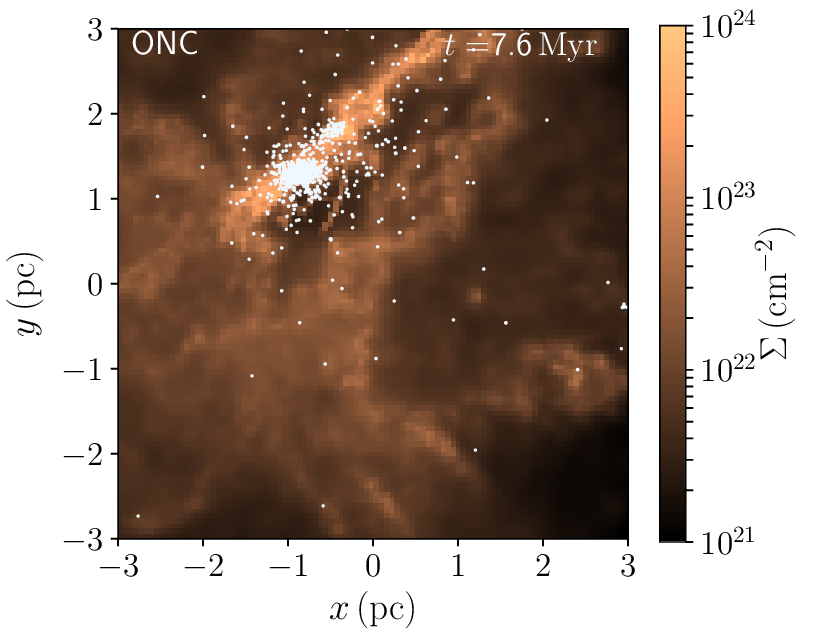}\\
    \caption{Snapshots at $t=5.8$, 6.3, 6.7, and 7.6\,Myr. }
    \label{fig:snapshots_3pc}
\end{figure*}

The distribution of star forming regions in the turbulent molecular clouds is clumpy.
In Fig.~\ref{fig:snapshots_3pc}, we present the snapshots of the central 6 pc $\times$ 6 pc region of the cloud, in which some clumps of stars are formed. In this simulation, all the clumps merged to the most massive clump (i.e., the main cluster). 

We detected the clumps using \textsc{hop} \citep{1998ApJ...498..137E} implemented in \textsc{amuse}. \textsc{hop} is a clump finding algorithm based on a friend-of-friend method using density peaks. We performed the clump finding in each snapshot and followed the merger history. Here, we detected clumps with more than 100 particles. We set the threshold densities for the cluster outer boundary ($\delta_{\rm outer}$) and density-peak detection ($\delta_{\rm peak}$) to be $1$ and $10 M_{\odot}$\,pc$^{-3}$, respectively \citep[see][for the details]{2019MNRAS.486.3019F}. 

\begin{figure}
	\includegraphics[width=0.95\columnwidth]{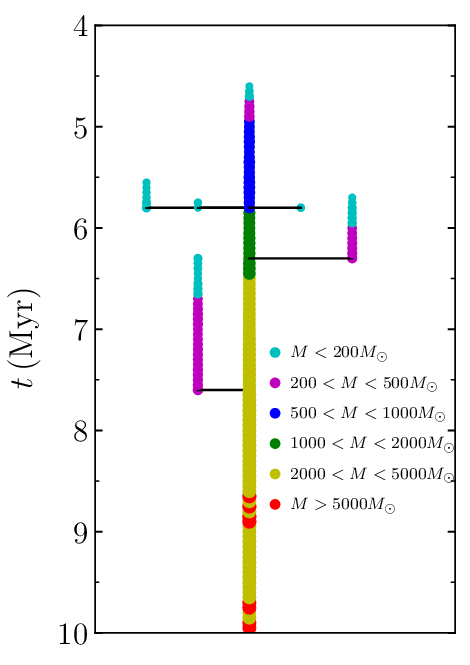}
	\caption{Merger tree of the simulated cluster. Colors indicate the cluster mass. Mergers occurred at 5.85, 6.35, and 7.65\,Myr.}
    \label{fig:merger_his}
\end{figure}

In Fig.~\ref{fig:merger_his}, we present the merger tree of the most massive star cluster found at 10\,Myr. 
The cluster experienced three mergers at 5.85, 6.35, and 7.65\,Myr. At 5.85\,Myr, three clumps merged to the main clump. The relative velocity before the merger was 3--5\,km\,s$^{-1}$. We note that our snapshots are stored every 0.05\,Myr, so we cannot resolve the mergers that occurred on a shorter timescale. This formation process of star clusters through clump mergers is crucial for the dynamical evolution of star clusters  \citep{2012ApJ...753...85F,2015PASJ...67...59F}. 

The final cluster mass reached $5000\,M_{\odot}$. The mass fraction of stars that is brought to the cluster by mergers is small. Most stars in the main cluster formed inside a cluster. We note that the cluster mass in our simulation is more than twice as massive as the ONC \citep[1800--2700\,$M_{\odot}$;][]{1998ApJ...492..540H}. The mass fluctuation at around 9 and 10\,Myr is related with the clump finding algorithm we adopted. The outer boundary of the clumps can slightly change in each snapshot. 

Once multiple massive stars are formed in the clusters, few-body encounters drive the ejections of massive stars from the cluster. Such ejected massive stars form H{\sc ii} regions at the outside of the cluster, in which the gas density is lower than the cluster centre (see the top panels of Fig.~\ref{fig:snapshots}). Even after the ionization of gas in the outer region, star formation continues in the cluster centre, in which dense gas still remains. As the star formation proceeds, the number of massive stars in the cluster increases, and dense cold gas is consumed. When the feedback energy from the massive stars is sufficient  to fully ionize the gas in the cluster, the star formation is quenched (see the bottom panel of Fig.~\ref{fig:snapshots}).

\subsection{Stellar age distribution}

\begin{figure}
	\includegraphics[width=0.95\columnwidth]{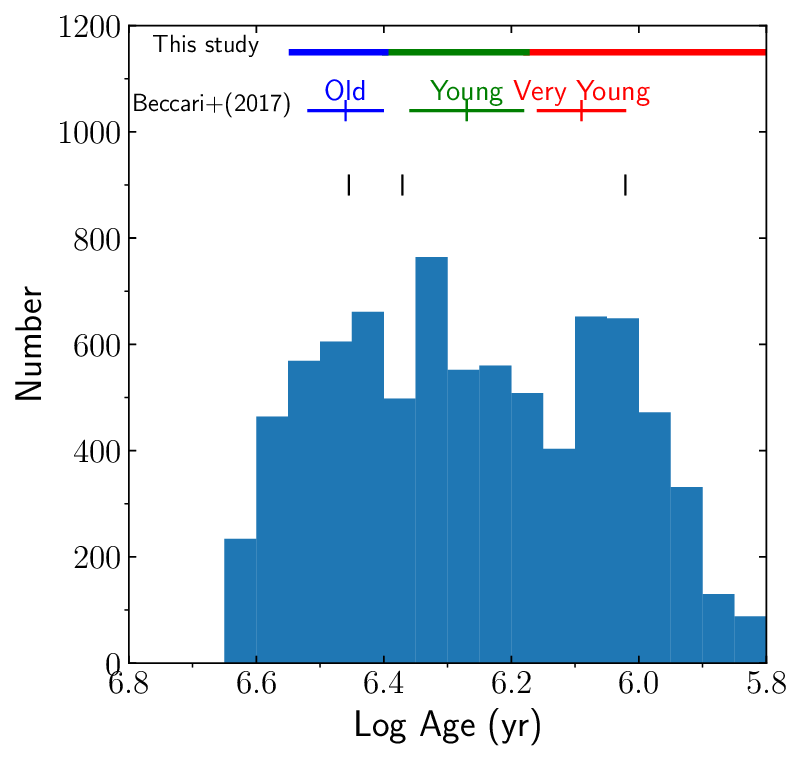}
	\caption{Stellar age distribution within 3\,pc of the cluster centre at $t=8.7$\,Myr. Black lines indicate the time of cluster mergers. Mergers occurred at 5.85, 6.35, and 7.65\,Myr.}
    \label{fig:stellar_age}
\end{figure}

We compare the stellar age distribution of stars in the simulation to the observation of the ONC.
Hereafter, we assumed the cluster at $t=8.4$\,Myr as an ONC-like cluster and observe the detailed structures of the cluster. At this time, the age distribution of stars in the main cluster is similar to that of the ONC \citep{2017A&A...604A..22B}; that is the age distribution shows three peaks. In Fig.~\ref{fig:stellar_age}, we present the age distribution of stars within 3\,pc from the cluster centre at this time. The cluster centre is determined using the density centre \citep{1985ApJ...298...80C} of stars in the main cluster. We compared this plot with Figure 6 of \citet{2017A&A...604A..22B}. We found three peaks of the stellar age in this distribution, which are similar to the three populations of the ONC, namely ``Very Young'', ``Young'', and ,``Old'', which are 1.08--1.53, 1.55--2.29, and 2.51--3.28\,Myr, respectively \citep{2017A&A...604A..22B}.
In this paper, we defined the three stellar populations with age ranges of $0.5<t_{\rm age}<1.5$, $1.5<t_{\rm age}<2.5$, and $2.5<t_{\rm age}<3.5$\,Myr as, ``Very Young'', ``Young'', and ``Old''.  These three periods include the three age-distribution peaks in the simulation. We summarize the definitions in Table~\ref{tb:Age}.

\begin{figure}
	\includegraphics[width=0.95\columnwidth]{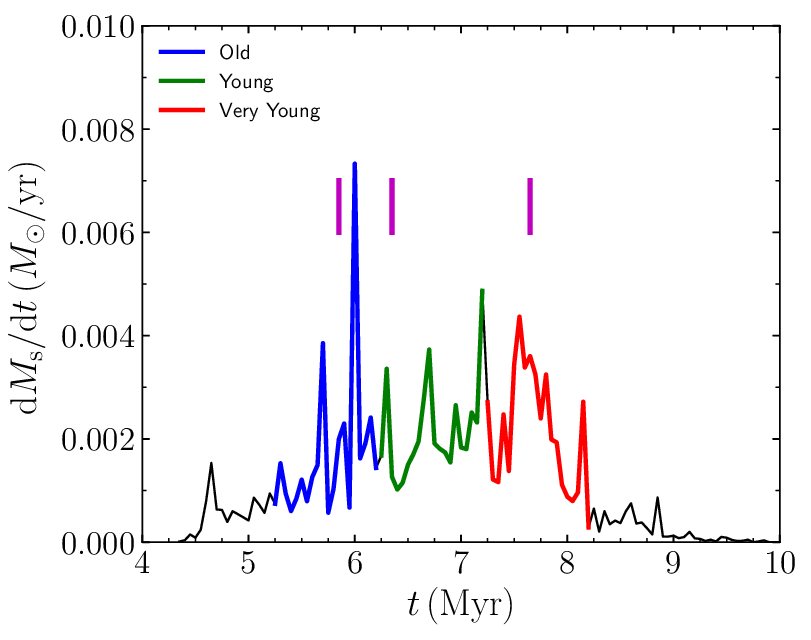}
	\caption{Star formation rate as a function of time. Colors are the same as Fig.~\ref{fig:total_stellar_mass}. Magenta lines indicate the timings of cluster mergers. The star formation rate is calculated using the snapshots, which are stored every 0.05\,Myr.}
    \label{fig:star_formation_rate}
\end{figure}

\begin{figure}
	\includegraphics[width=0.95\columnwidth]{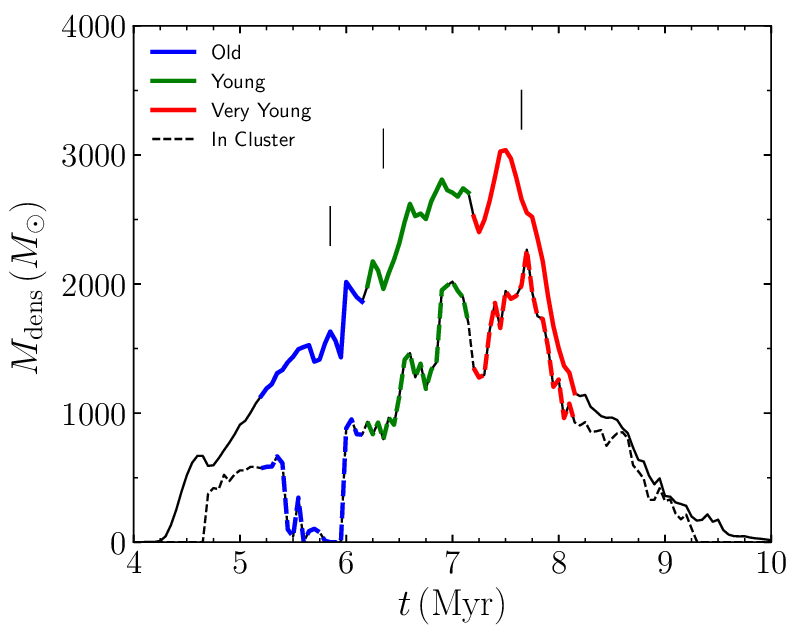}
	\caption{Dense ($>10^5$\,cm$^{-3}$) cold ($20$\,K) gas mass as a function of time. Full and dashed curves indicate the total and 0.3\,pc from the main cluster centre, respectively. Black vertical lines indicate the timings of cluster mergers. }
    \label{fig:dens_gas_mass}
\end{figure}

In Fig.~\ref{fig:stellar_age}, we also present the time at which clump mergers occurred. The three peaks roughly correspond to the mergers.
To understand the relation between the clump mergers and the star formation, we investigated the star formation rate as a function of time. The result is shown in Fig.~\ref{fig:star_formation_rate}. There are several peaks in the star formation rate, and some of them seem to be coincident with the clump mergers. As the star formation rate should be correlated with the amount of dense cold gas, we also investigate the mass of dense ($n_{\rm H}>10^5$\,cm$^{-3}$) and cold ( $T=20$\,K) gas in the simulation. In Fig.~\ref{fig:dens_gas_mass}, we present the total mass of dense cold gas for the entire system and within 0.3\,pc from the centre of the main cluster. The peaks of the gas mass in the cluster centre increase soon after the mergers. 

We note that our method to determine the density centre sometimes fails when some clumps are merging because the "clump" detected using the clump finding method may include sub-clumps in the region. When two or more clumps are included in the given stellar distribution, the method determines the middle of two clumps as the density centre, which happened at $t \sim 5.5$--6\,Myr. This is the reason why there are no dense gas in the cluster centre in this period. As shown in Fig.~\ref{fig:snapshots_3pc}, the clumps were in a pre-merger state in this period.

As shown in Fig.~\ref{fig:snapshots_3pc}, the clumps formed in cold-gas streams accrete to the main cluster with the gas. 
When clumps merge to the main cluster, they bring cold gas into it. The potential of the main cluster is deepened by the mergers, and hence the amount of dense gas increases in the cluster centre. This process enhances the star formation in the cluster.

\begin{figure*}
	\includegraphics[width=0.65\columnwidth]{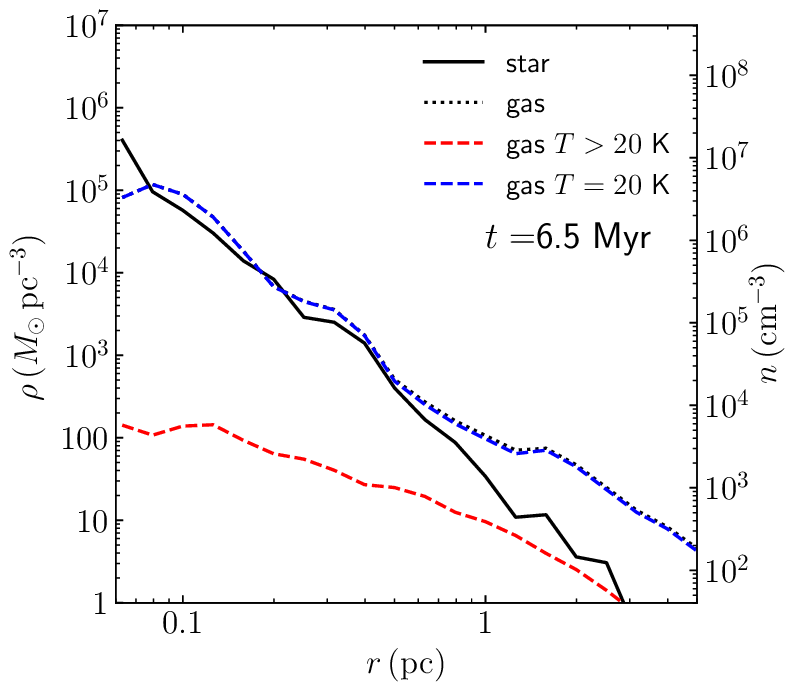}
	\includegraphics[width=0.65\columnwidth]{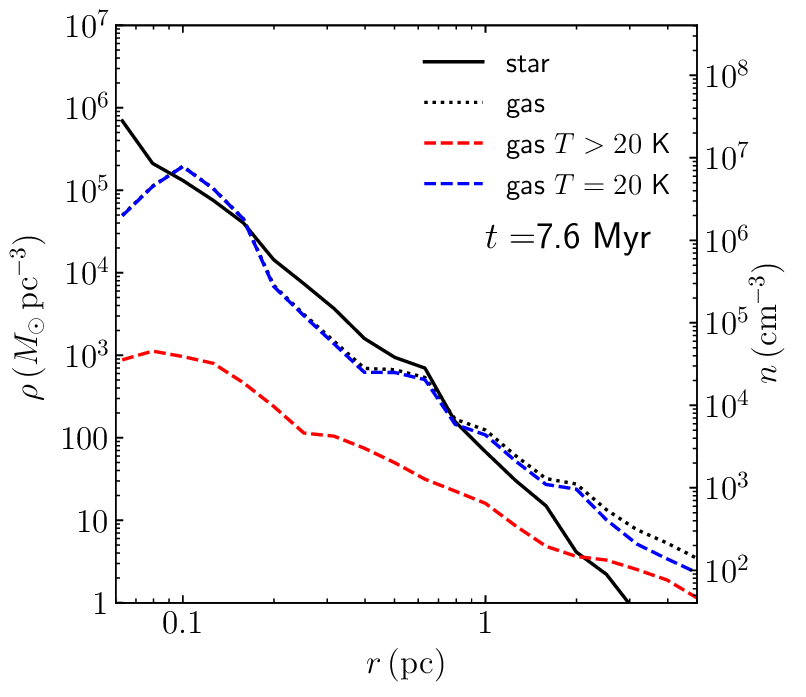}
	\includegraphics[width=0.65\columnwidth]{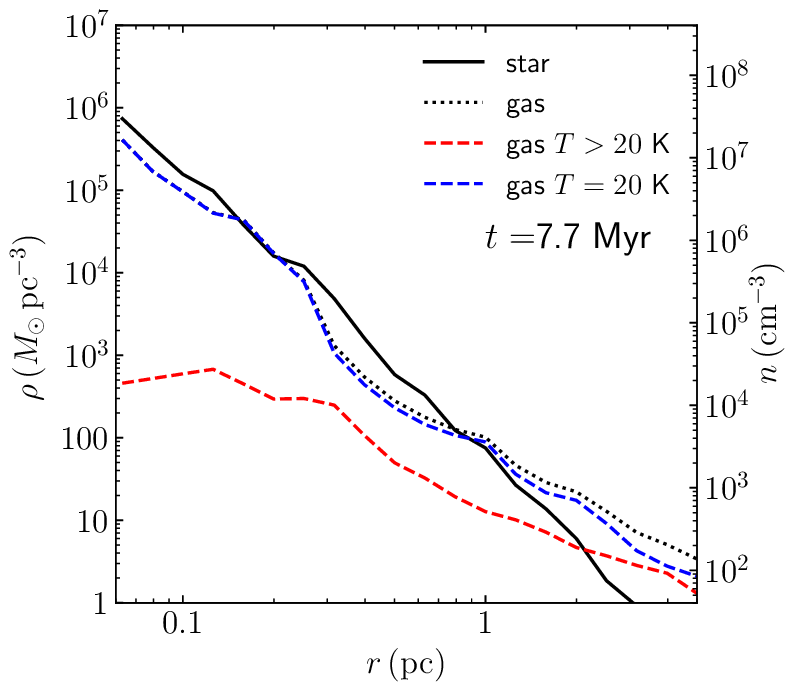}\\
	\includegraphics[width=0.65\columnwidth]{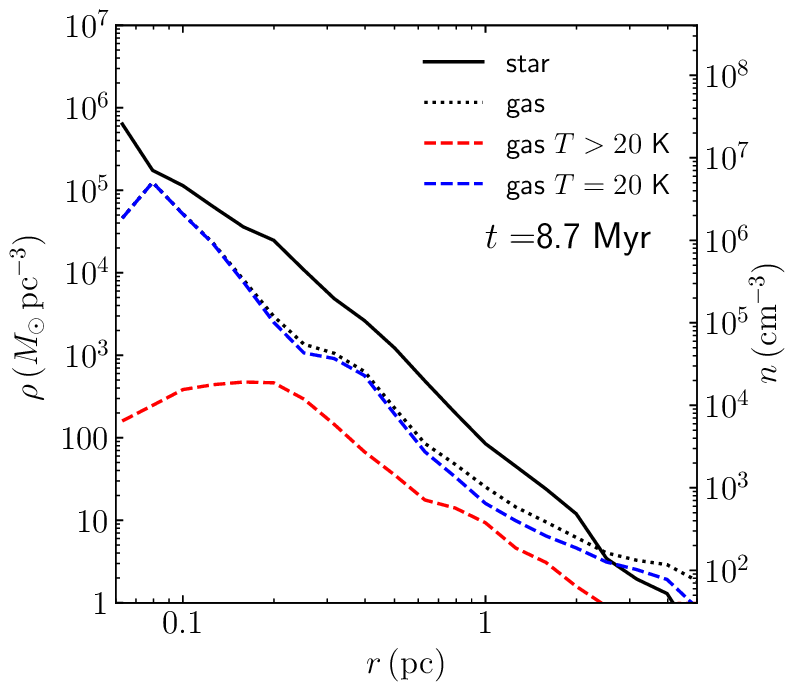}
	\includegraphics[width=0.65\columnwidth]{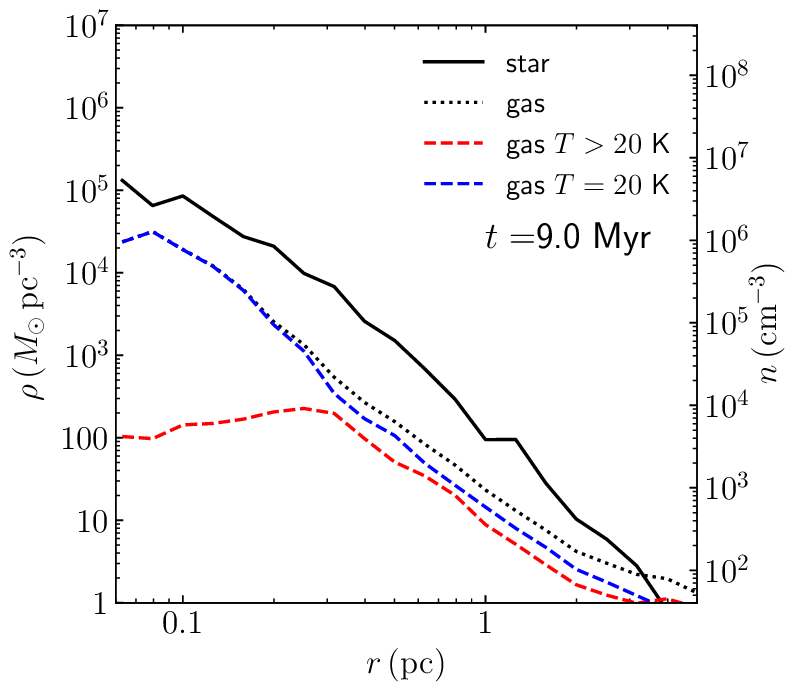}
	\includegraphics[width=0.65\columnwidth]{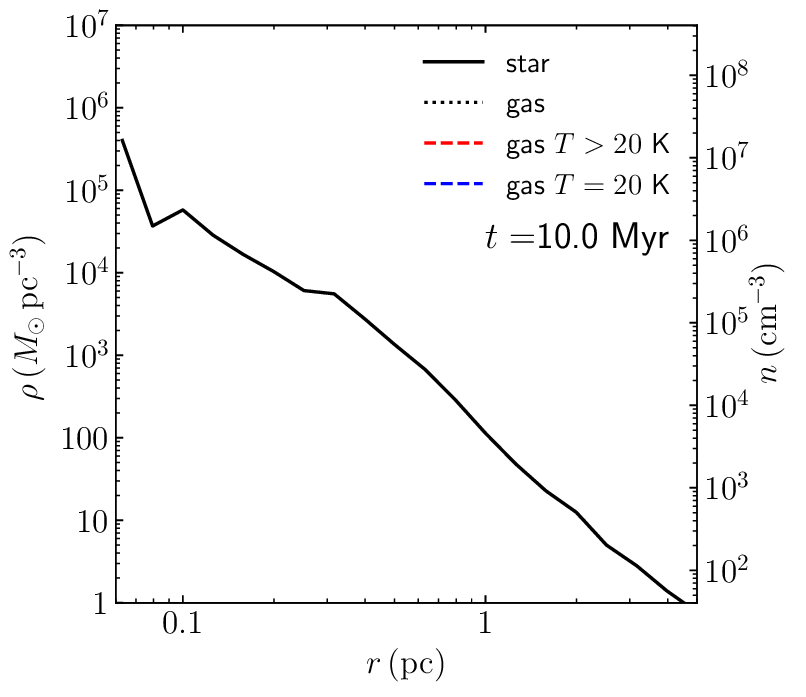}\\
	\caption{Stellar and gas density profiles of the main cluster.}
    \label{fig:star_gas_prof}
\end{figure*}

The increase of the dense cold gas in the cluster centre is confirmed by the density profiles of the cluster. 
In the top panels of Fig.~\ref{fig:star_gas_prof}, we present the gas and stellar density profiles of the main cluster for $t=7.6$--$7.7$\,Myr. A merger occurred at around $7.65$\,Myr. Before the merger ($t=7.6$ and 7.65\,Myr), the cold gas in the cluster centre is depleted due to the feedback from the massive stars in the cluster. After the merger of a clump ($t=7.7$\,Myr), however, the density of cold gas increases in the cluster centre. This results in the formation of stars inside the main cluster.

At $t\sim 7$\,Myr, there is a peak of the dense cold gas in the cluster. Although we did not identify any mergers at this time (see Fig.~\ref{fig:merger_his}), clumps smaller than our detection limit (100 stars) accreted to the main clump at around this time. In the bottom left panel of Fig.~\ref{fig:snapshots_3pc}, we present the snapshot at $t=6.7$\,Myr. There is a clump (left bottom from the main cluster) orbiting around the main cluster. In the left direction from the main cluster, there are some small clumps merging to the main cluster. The accretion of these small clumps also caused the star formation at around $t=7$\,Myr. 

The timing of clump mergers depends on the random seeds for the initial turbulent velocity fields and the randomness of the formation of massive stars. Therefore, the initial conditions with the same parameters do not always result in the same formation history due to such random effects that we allowed. The final cluster mass and the numbers of clusters also largely changes \citep[see ][]{2021PASJ...73.1057F,2021PASJ...73.1074F}. In this paper, we present the result indicating  star formation peaks similar to the ONC, but the other runs starting from the same initial mass, density, and virial ratio,  with different random seeds for the turbulence, did not show the three peaks. Thus, the shape of the stellar age distribution is not deterministic but random.

We briefly discuss the binaries in our simulation. As discussed in \citet{2021PASJ...73.1057F}, the number of hard binaries in our simulation is much smaller than observed ones. This is because our simulation cannot resolve the physical scale for hard binary formation (i.e., 100-au scale). There is only one binary in the main cluster, which is dynamically formed in the cluster via binary hardening due to repeating binary-single encounters. To address the discrepancy between the simulation and observation, we should assume the formation of primordial binaries when we create star particles. With such a treatment, we may observe complete ejections of massive stars, as presented  in \citet{2019MNRAS.484.1843W}.

\subsection{Triaxial structure of the star cluster}

Observationally, anisotropic velocity distributions are found in the central 0.5\,pc of the ONC \citep{2019AJ....157..109K,2019ApJ...870...32K,2022ApJ...926..141T}. 
We investigated the effect of clump mergers on the velocity anisotropy through our simulation.

\begin{figure}
	\includegraphics[width=0.95\columnwidth]{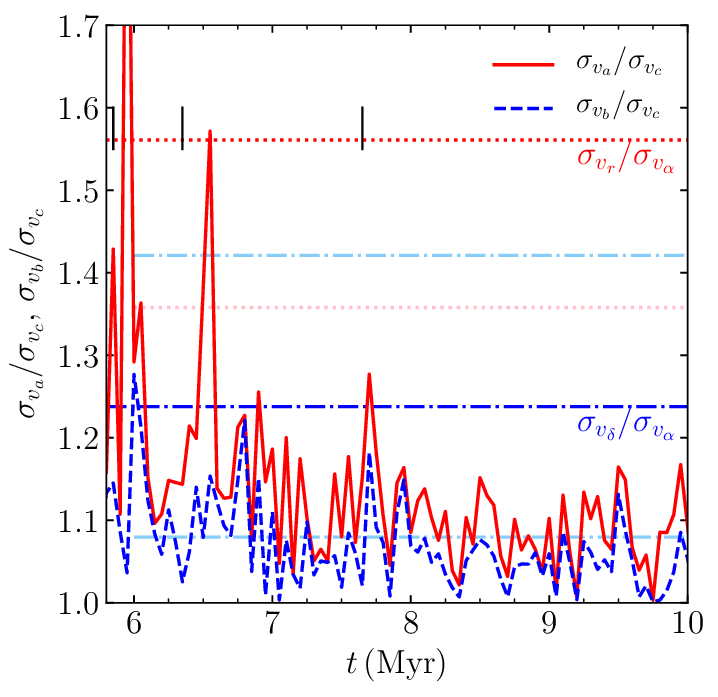}
	\caption{Velocity dispersion ratios of stars ($0.5<m<1\,M_{\odot}$) in the inner 0.5\,pc of the simulated cluster as a function of time. Red full and blue dashed curves indicate the velocity dispersion ratios of the simulated cluster, and red dotted and blue dash-dotted lines indicate the observed values \citep{2022ApJ...926..141T}. Pink dotted and light-blue dash-dotted lines are the errors in the observation, and vertical lines indicate the timings of clump mergers. }
    \label{fig:vel_aniso}
\end{figure}

In each snapshot, we calculated the principal axes of inertia for the velocities of the stars in the inner 0.5\,pc from the cluster centre. We used stars with masses of $0.5<m<1\,M_{\odot}$ because the samples in \citet{2022ApJ...926..141T} are low-mass stars. Then, we measured the velocity dispersions of the stars along with the principal axes. We defined the largest, middle, and lowest velocity dispersions as $\sigma_{v_{a}}$, $\sigma_{v_b}$, and $\sigma_{v_c}$, respectively.
In Fig.~\ref{fig:vel_aniso}, we present the time evolution of the velocity dispersion ratios. We calculated the velocity dispersions later than 6\,Myr, at which the main clump includes  enough stars within 0.5\,pc. For comparison, we also plotted the observed velocity dispersion ratios obtained from \citet{2022ApJ...926..141T}. As our simulated cluster is more massive than the ONC, we compared the velocity dispersion ratio, rather than the velocity dipsersions themselves.

As shown in the figure, the velocity dispersion ratios increase after mergers, but decrease in $\sim 0.5$\,Myr. According to \citet{2022ApJ...926..141T}, the largest velocity dispersion ratio is $\sigma_{v_{r}}/\sigma_{v_{\alpha}}=1.44^{+0.21}_{-0.18}$. The time at which $\sigma_{v_a}/\sigma_{v_c}$ reaches this range is limited to the moment just $<0.1$\,Myr after a merger. The condition of $\sigma_{v_{\delta}}/\sigma_{v_{\alpha}}=1.23^{+0.17}_{-0.16}$ is satisfied for $\sim 0.5$\,Myr after a merger. As the dynamical time of this cluster is calculated to be $\sim 0.2$\,Myr, assuming that $5000 M_{\odot}$ within 1\,pc, the  anisotropy disappears in a few dynamical times.  

Later than $t=9$\,Myr, the system is almost gas-free, and no mergers occurred.  Even in this quiet period, the velocity dispersion ratio fluctuated at around 1.1. It seems that a velocity dispersion ratio smaller than 1.2 is a dynamically steady state. If we measure a velocity dispersion ratio more than $\sim 1.2$, it implies a recent clump merger. 
From other studies on the velocity measurement of the ONC centre, the velocity dispersion rate of $1.35\pm0.07$ \citep{2019AJ....157..109K} and $1.6\pm0.4$ \citep{2019ApJ...870...32K} are obtained. Thus, the ONC is considered to have experienced a clump merger within the past $\sim 0.5$\,Myr.

\subsection{Virial ratio}
The virial ratio (kinetic over potential energy) is another parameter to measure the dynamical states of stellar systems.
We also calculated the time evolution of the virial ratio of the main cluster. We measured the virial ratio of stars within 0.5\,pc from the density centre of the main cluster. The cluster mean velocity was determined by the centre-of-mass velocity of stars within 0.5\,pc. We calculated the total kinetic energy excluding gas, but the potential energy of stars includes the potential due to the gas. If the system is in virial equilibrium, the virial ratio is 0.5.  

\begin{figure}
	\includegraphics[width=0.95\columnwidth]{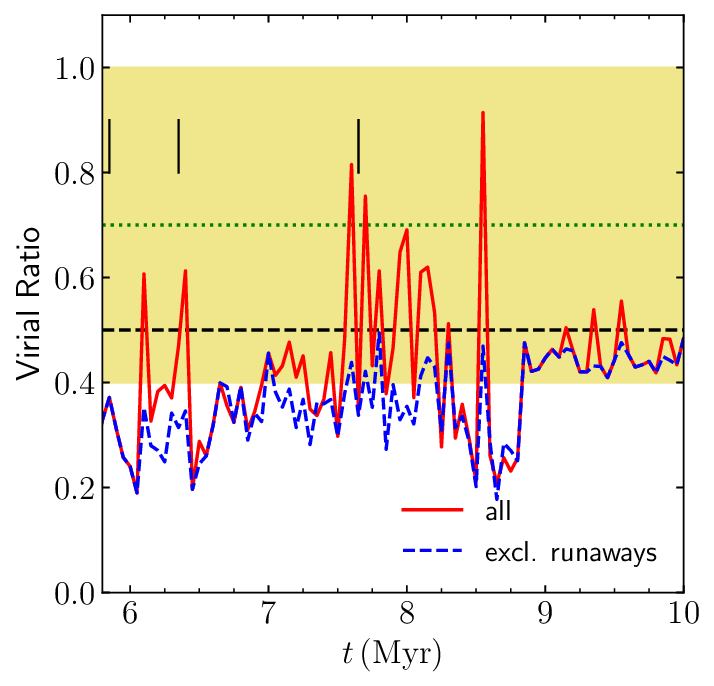}
	\caption{Virial ratio (kinetic over potential energy) of the stars within 0.5\,pc of the cluster centre for all stars (red full) and excluding runaways (blue dashed). Black horizontal dashed line indicate "virialized" (0.5). Black vertical lines indicate the timings of clump mergers. Green dashed line and yellow region indicates the observed value with an error ($0.7\pm0.3$).}
    \label{fig:virial_ratio}
\end{figure}

The time evolution of the virial ratio is shown in Fig.~\ref{fig:virial_ratio}. Once a merger occurred, the virial ratio increased. After the mergers at $t=5.85$ and $6.35$\,Myr, the virial ratio immediately decreased to sub-virial. The decrease of the virial ratio is due to the existence of dense cold gas in the cluster. The cold gas collapsed to the cluster centre, and the dynamically cold gas formed new stars. As a result, the new born stars are also dynamically cold. Therefore, the cluster embedded in cold gas is sub-virial. During $t=6.5$--7.5\,Myr, the period between the second and the third mergers, the cluster centre maintained a sub-virial state (see Fig.~\ref{fig:virial_ratio}). 

After the merger at $t=7.65$\,Myr, the virial ratio increased again, but this time it decreased more slowly compared with the merger at $t=6.35$\,Myr. As shown in the density profile in Fig.~\ref{fig:star_gas_prof}, the stellar mass in the cluster centre exceeded the gas mass at $t=7.7$\,Myr. Around $\sim 1$\,Myr after the merger, the cluster centre became sub-virial again. 

\begin{figure}
	\includegraphics[width=0.95\columnwidth]{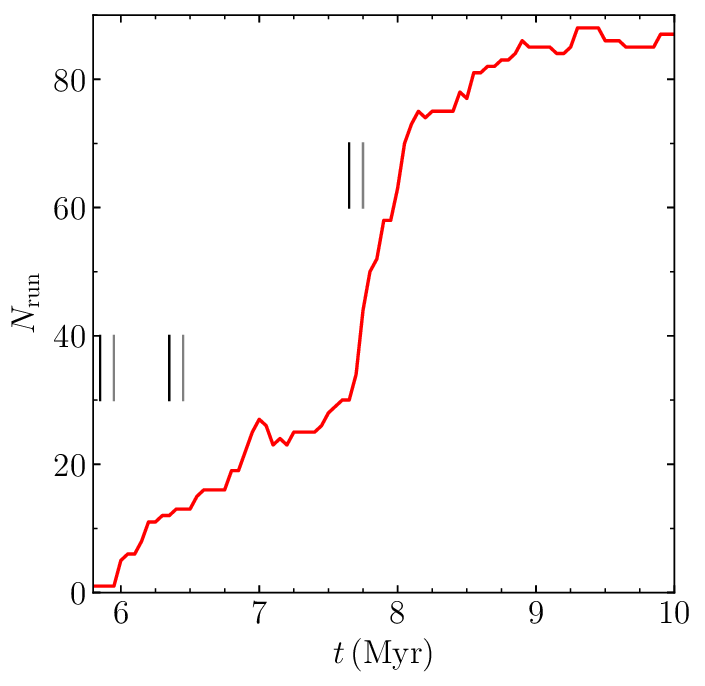}
	\caption{Number of runaway stars located at $>3$\,pc from the cluster centre as a function of time. Black lines indicate the merger time, and gray lines indicate merger time $+0.1$\,Myr, which is the time for stars with 30\,km\,s$^{-1}$ to reach 3\,pc.}
    \label{fig:N_runaways}
\end{figure}

We found that the super-virial state is caused by the formation of runaway stars, which are stars with a velocity of 30\,km\,s$^{-1}$. In Fig.~\ref{fig:N_runaways}, we present the number of runaway stars that are located farther than 3\,pc from the cluster centre. The number of runaways jumps up after clump mergers. Specifically,  after the merger at $t=7.65$\,Myr, with around half of the runaway stars being formed in this period. The peak at around $t=7$\,Myr is related to the dense gas peak in the cluster caused by minor mergers (see Fig.~\ref{fig:dens_gas_mass}). We also calculated the virial ratio excluding runaway stars (see Fig.~\ref{fig:virial_ratio}) and confirmed that the increase of the virial ratio is caused by the existence of runaways. The virial ratio also increased at $t=8.55$\,Myr. At this time, two fast runaway stars with $>100$\,km\,s$^{-1}$ existed.

At around 9\,Myr, the dense gas in the cluster was completely blown away, and the cluster became gas-free (see Fig.~\ref{fig:dens_gas_mass} and the bottom panels in Fig.~\ref{fig:star_gas_prof}). We confirmed that the cluster was virialized after gas expulsion (see Fig.~\ref{fig:virial_ratio}).

We also compare our results with the measurements for the ONC. 
The virial ratio of the ONC centre ($<0.5$\,pc) is estimated to be $0.7\pm0.3$ \citep{2019AJ....157..109K}. Including the measurement error, the virial ratio obtained in our simulation is in this range for most of the time (see Fig.~\ref{fig:virial_ratio}). However, if we assume that the ONC centre is virial or slightly super-virial and that the gas mass is comparable to the stellar mass \citep{2006ApJ...641L.121T}, the virial ratio after the merger at $t=7.65$\,Myr suggests that the ONC experienced a clump merger within the past 1\,Myr.

\subsection{Runaway stars}
\begin{figure}
	\includegraphics[width=0.95\columnwidth]{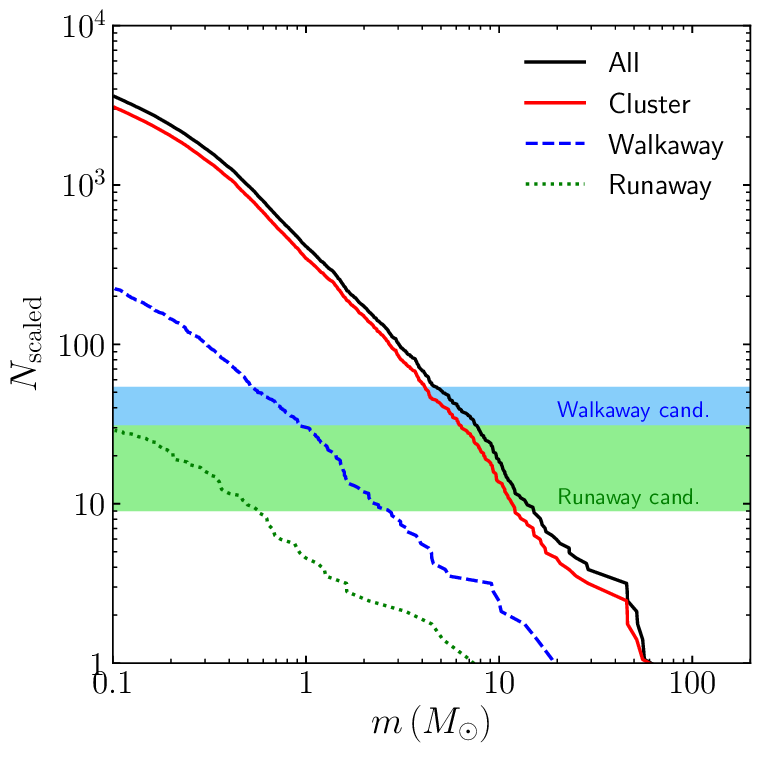}
	\caption{Mass functions of all stars, stars in the star cluster (within 3\,pc from the cluster centre), runaways, and walkaways. Green and blue shaded regions are the number of runaway and walkaway candidates in observations.}
    \label{fig:runaway_mass}
\end{figure}

As mentioned above, a number of runaway stars are formed in our simulation. 
We find 82 runaway ($v>30$\,km\,s$^{-1}$) and 248 walkaway ($10<v<30$\,km\,s$^{-1}$) stars further than 3\,pc from the cluster centre at $t=8.7$\,Myr.  
In Fig.~\ref{fig:runaway_mass}, we present the cumulative mass functions of runaway and walkaway stars, but the number is scaled to the mass of the ONC because our simulated cluster is more massive than the ONC.
We also present mass functions of the all stars in the simulation and stars in the cluster (within 3\,pc from the cluster centre). We confirm that the entire mass function follows the mass function that we assumed (Kroupa mass function), and the mass function inside the cluster also follows the given one. 
The power of the mass functions for runaways and walkaways are shallower than the given one. This is because more massive stars in the cluster centre are more efficiently ejected due to the three-body encounters \citep{2011Sci...334.1380F}.

In Fig.~\ref{fig:runaway_mass}, we also show the numbers of observed runaway and walkaway candidates \citep{2019ApJ...884....6M,2020MNRAS.495.3104S,2020AJ....159..272P}. We can conclude that most runaway stars formed by the ONC have already been found. On the other hand, more than 100 walkaways may have not been found yet. However, most of them are low-mass stars, which are hardly observed. 

In addition, \citet{2020MNRAS.495.3104S} reported that all runaways and walkaways are less massive than $8\,M_{\odot}$. In our simulation, the most massive runaway and walkaway stars are $\sim 7\,M_{\odot}$ and $\sim 20\,M_{\odot}$, respectively. Since the most massive stars in the cluster remain in the cluster centre, the most massive stars among runaways and walkaways are less massive compared to the stars in the cluster.

\begin{figure}
	\includegraphics[width=0.95\columnwidth]{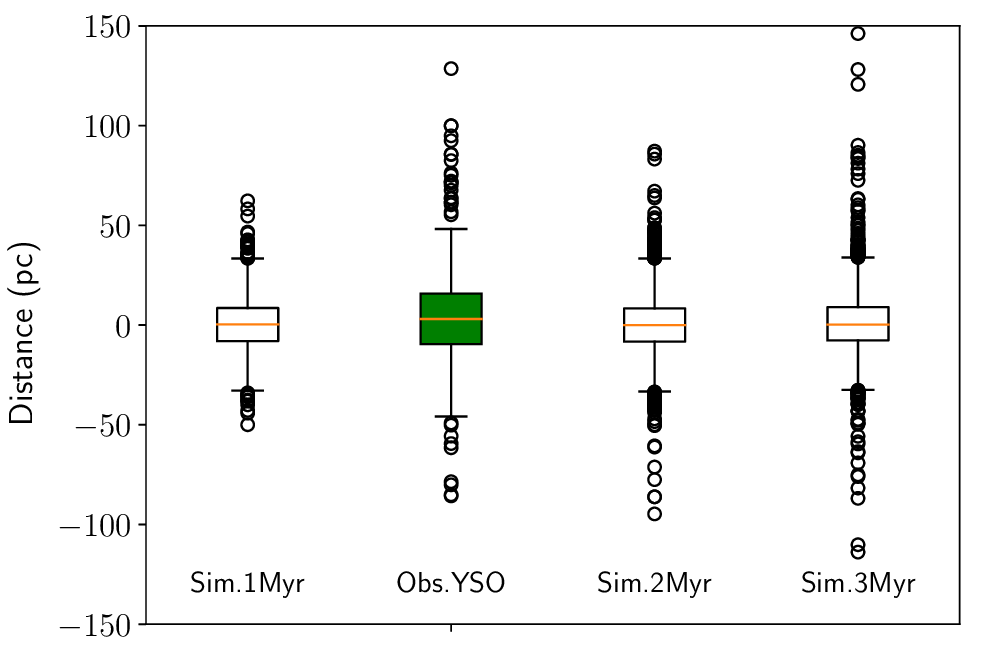}
	\caption{The radial distribution of the stars in the simulation and observation. The green box plot represents the observed YSOs \citep{2018A&A...619A.106G}.}
    \label{fig:YSO_dist}
\end{figure}

This ejection of stars result in a wide spatial distribution of young stars. In Fig.~\ref{fig:YSO_dist}, we present the radial distribution of stars in each age bin. We add errors randomly chosen from a Gaussian function with a dispersion of $0.08$ to the parallax of the simulation data. We assume that the distance to the ONC as 390\,pc \citep{2017ApJ...834..142K}. 
We find that, even the youngest stars ($<1$\,Myr), they can reach more than 50\,pc from the cluster. 
Stars older than 2\,Myr reach 100\,pc. In the same figure, we also present the distance distribution of young stellar objects (YSOs).

\subsection{Radial age gradient of the ONC}

We also compared the radial distribution of our simulated cluster to that of the ONC. 
The bottom left panel of Fig.~\ref{fig:star_gas_prof} shows the gas and stellar density distribution of the cluster at $t=8.7$\,Myr. The cluster is already dominated by stars. However, the gas mass of the ONC is estimated to be comparable to the stellar mass \citep{2006ApJ...641L.121T}. From this point of view, earlier time in the simulation (e.g., $t=7.7$\,Myr) may be closer to the ONC. 

\begin{figure}
	\includegraphics[width=0.95\columnwidth]{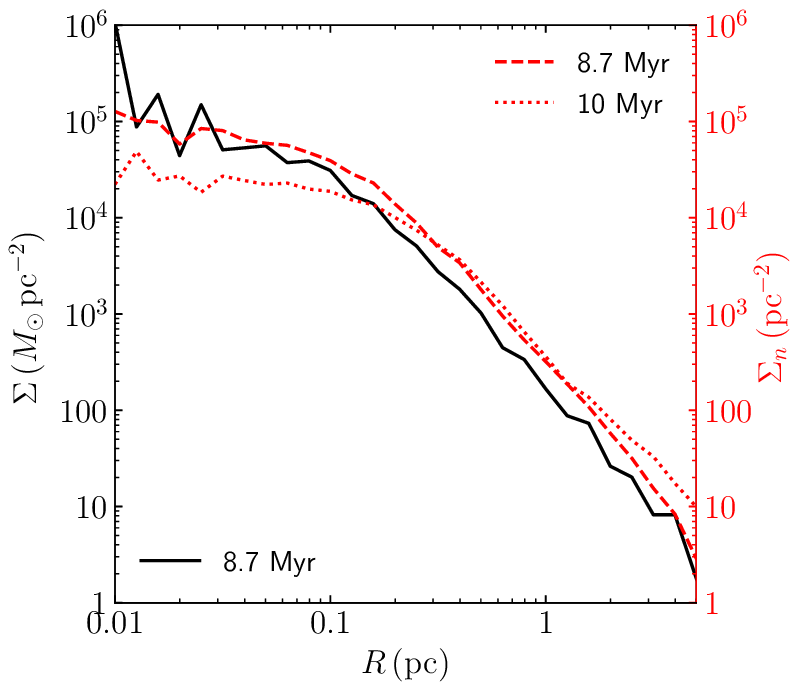}
	\caption{Surface density profile of stars at $t=8.7$\,Myr (black curve) and surface number densities of the cluster at $t=8.7$\,Myr (rad dashed) and $t=10$\,Myr (red dashed).}
    \label{fig:prof_2d}
\end{figure}

In Fig.~\ref{fig:prof_2d}, we present the surface density profile of stars at $t=8.7$\,Myr. In the cluster centre, the highest density reaches $10^6M_{\odot}$\,pc$^{-2}$. The surface density of the ONC is $\sim 3\times 10^5\,M_{\odot}$\,pc$^{-2}$ at $R=0.01$\,pc \citep{2014ApJ...795...55D}. Considering the difference in the cluster masses, the central densities are comparable. The central high density is caused by the concentration of massive stars in the cluster core due to the mass segregation. We also present the surface number density profile in this figure. The number density shows a core halo profile.

We also calculated the three-dimensional core radius and density using the method developed by \citet{1985ApJ...298...80C}.
The three-dimensional core radius and density are 0.05 pc and $9.3\times 10^5 M_{\odot}\,{\rm pc}^{-3}$, respectively. The three-dimensional half-mass radius is 0.33 pc, and the half-mass density is $1.9\times 10^4 M_{\odot}\,{\rm pc}^{-3}$ for the stars determined as cluster members through HOP.
The core radius of the ONC was measured to be 0.20\,pc for the surface density \citep{1998ApJ...492..540H}. However, recent study by \citet{2014ApJ...795...55D} suggested that the density profile continues as a power-law function rather than a flat core. With their model, the density within 0.05\,pc is $1.8\times10^5 M_{\odot}\,{\rm pc}^{-3}$. We note that the core density and core radius in the simulation highly fluctuates for snapshot by snapshot.

After the complete gas removal ($t=10$\,Myr, see the bottom right panel of Fig.~\ref{fig:star_gas_prof}), the bound cluster mass was $3925\,M_{\odot}$. At this time, the half-mass radius and density were $0.42$\,pc and $6.2\times 10^{3} M_{\odot}$\,pc$^{-3}$, respectively. We also present the surface number density of the cluster at $t=10$\,Myr in Fig.~\ref{fig:prof_2d}. The central surface density of the cluster decreased to one third of that at $t=8.7$\,Myr. Thus, the central density of the cluster decreases after the gas expulsion, but this cluster survived the gas expulsion.

\begin{figure*}
	\includegraphics[width=0.65\columnwidth]{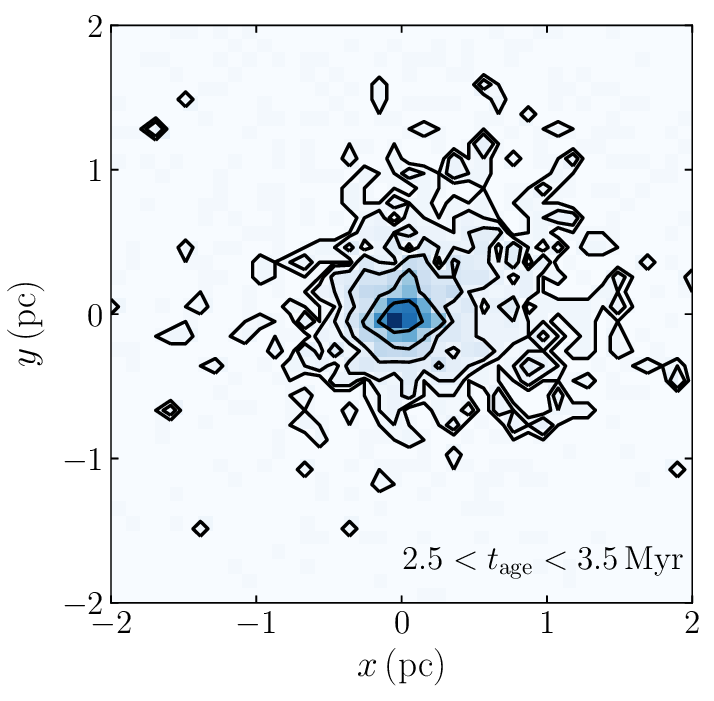}
	\includegraphics[width=0.65\columnwidth]{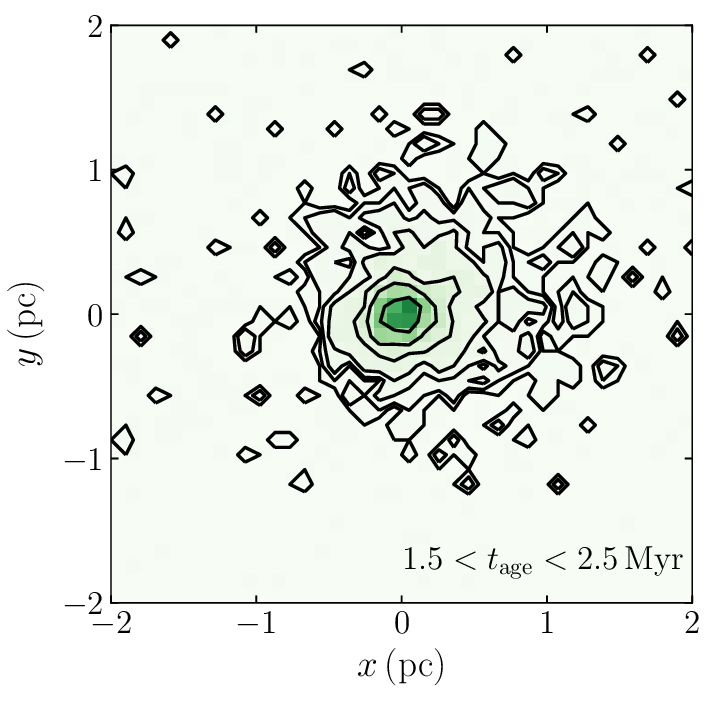}
	\includegraphics[width=0.65\columnwidth]{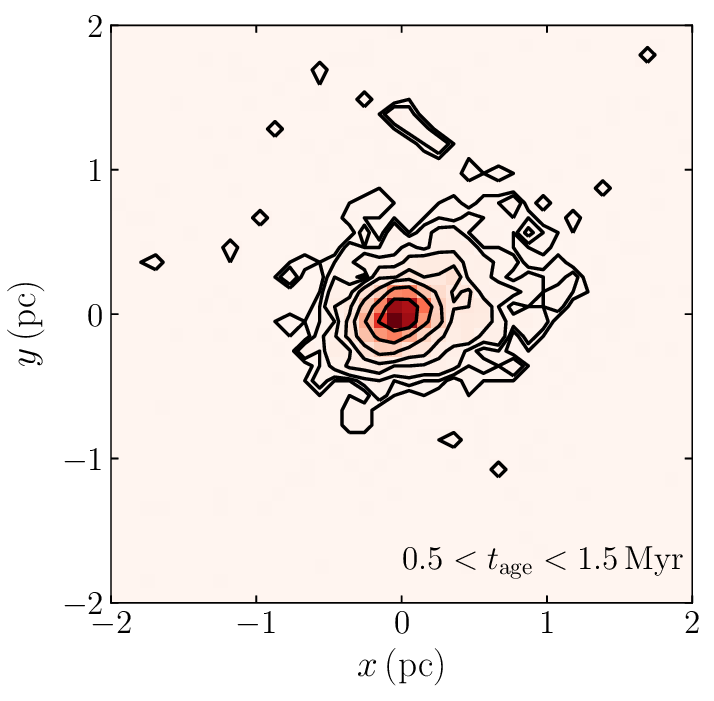}
	\caption{The distribution of stars in each age range $t=8.7$\,Myr.}
    \label{fig:stellar_age_dist}
\end{figure*}

\begin{figure}
	\includegraphics[width=0.95\columnwidth]{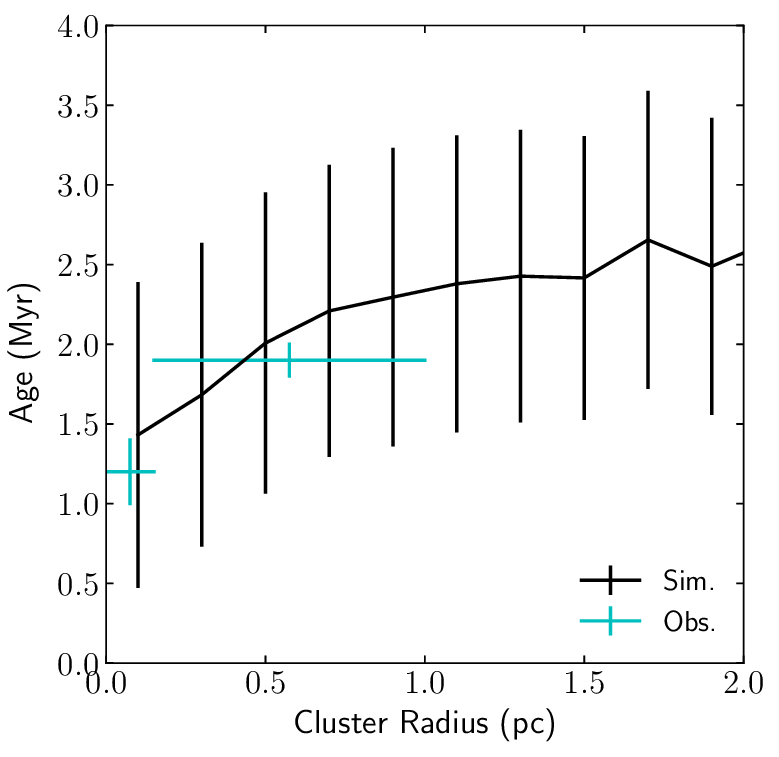}
	\caption{Radial distribution of the mean stellar age  at $t=8.7$\,Myr (black curve). The horizontal axis shows the projected distance from the cluster centre. The vertical line indicate the standard deviation. Cyan points shows the observed stellar age in the ONC; $1.2\pm0.2$\,Myr ($R<0.15$\,pc) and $1.9\pm0.1$\,Myr ($0.15<R<1$\,pc) \citep{2014ApJ...787..109G}}
    \label{fig:age_rad}
\end{figure}

In Fig.~\ref{fig:stellar_age_dist}, we present the spatial distribution of ``Very Young'', ``Young'',  and ``Old'' stars at $8.7$\,Myr. Similar to Fig.~3 of \citet{2017A&A...604A..22B}, ``Very Young'' stars are concentrated to the cluster centre. This is because the star formation continues in the cluster centre, at which dense cold gas remains until the last moment (see Fig.~\ref{fig:dens_gas_mass}). We also found  the radial gradient of stellar age in our simulated cluster. In Fig.~\ref{fig:age_rad}, we present the radial distribution of the mean stellar age. We found  the mean age increases from the cluster centre ($\sim 1.4$\,Myr) towards to the outer region of the cluster, as was also observed in the ONC \citep{2014ApJ...787..109G}. In the same plot, we also presented the mean age measured in the ONC; $1.2\pm0.2$\,Myr at $R<0.15$\,pc and $1.9\pm0.1$ at 0.15--1\,pc \citep{2014ApJ...787..109G}. Although the dispersions in the simulation are much larger than the observation, the observed cluster age in smaller annuli also has dispersions similar to the simulation \citep[see Fig.~1 in ][]{2014ApJ...787..109G}. Our simulation suggests that the age gradient observed in the ONC is caused by the star formation in the dense gas remaining in the cluster centre.

\section{Summary}
We performed an $N$-body/SPH simulation for the formation of a star cluster,
which resembles the ONC.
We compared the spatial and kinematic structures and the formation history of the cluster in our simulation with those of the ONC. 

We found that the main cluster formed in our simulation experienced three major mergers. In every mergers, dense cold gas was brought to the cluster, and it accreted to the cluster centre and triggered the star formation. This process resulted in the three peaks in the age distribution of stars in the formed cluster, suggesting that the ONC also experienced three mergers during its formation. 

We did not find complete ejection of massive stars from star clusters as was suggested by \cite{2018A&A...612A..74K} and \cite{2019MNRAS.484.1843W}, which can form multiple populations of stars without mergers of sub-clusters (ejection scenario). 
In our simulations, however, we cannot include initially tight binaries and multiples, which are common for massive stars \citep{2012Sci...337..444S}.
Therefore, we cannot directly test the ejection scenario.
To treat the formation of primordial tight binaries and multiples without using a stochastic sampling from a given IMF, we need a extremely high resolution simulation resolving an au scale, which prevent a simulation of a large scale star cluster formation such as gas flow on 100-pc scale. Alternatively, we may assume binary formation as a sub-grid physics.
Thus, the role of primordial binaries and multiples should be tested in future simulations.

The clump mergers also caused the anisotropy in the velocity dispersions of stars in the cluster. Although the anisotropy in the velocity dispersion disappeared within $\sim 0.5$\,Myr, the anisotropy was as strong as the observed one. The virial ratio was also increased by the clump mergers. The increase of the virial ratio is caused by the formation of runaway stars. Clump mergers enhance the formation of runaway stars, and the runaways contribute to the increase of the virial ratio. The observed virial or super-virial state of the ONC can be explained by the clump merger.
Thus, the kinematic measurements of the ONC imply that the ONC experienced a clump merger within the past 0.5\,Myr.

The numbers of runaway stars formed in our simulation is comparable to that of observed candidate. On the other hand, more than one hundred of walkaways may have not observationally been found yet. The three-body encounters eject young stars born in the central region of the star cluster. As a result, young stars distribute even 50--100\,pc away from the cluster. This can explain the wide distribution of YSOs in the Orion region. 

We also compared the radial distribution of the stellar age in our simulated cluster to the observations of the ONC. We found a stellar age gradient from the cluster centre to the outer region. The mean age in the inner 0.1\,pc in our simulation was $\sim 1.4$\,Myr. The mean age increased with increasing radius and reached to $\sim 2.5$\,Myr in the region further than 1\,pc from the cluster centre
This result is consistent with the observed one: $1.2\pm0.2$\,Myr ($<0.1$\,pc) and $1.9\pm0.1$\,Myr (0.15--1\,pc) \citep{2014ApJ...787..109G}. 

The radial age gradient was caused by  star formation in the cluster centre until the latest time. This is because the cold dense gas, which can form stars, remains in the cluster centre rather than the outer region. The gas density is low in the outer region of the cluster, and therefore the gas is easily ionized by massive stars ejected from the cluster centre as a result of binary-single encounters. Meanwhile, the gas density at the cluster centre is too high to be fully ionized. Therefore, dense gas remains in the cluster centre until the end and continues the star formation in the cluster core. 

Once enough massive stars are formed in the cluster and/or the dense gas is consumed by the star formation, the massive stars in the cluster fully ionize the gas. Then, all the gas in the cluster is blown away. The central density of the cluster decreased to one-third of that before the gas expulsion. Even after the gas expulsion, $\sim 2/3$ of stars remained as bound stars.

\section*{Acknowledgments}
The authors thank Steven Rieder for providing the \textsc{amuse} script for surface density and temperature map and Kurumi Ishikura and Ryoichi Nishi for discussion, Takaaki Takeda (4D2U at National Astronomical Observatory of Japan) for the visualization of the simulation, and Editage (www.editage.com) for English language editing. 
Numerical computations were carried out on Cray
XC50 CPU-cluster at the Center for Computational Astrophysics (CfCA) of the National Astronomical Observatory of Japan. 
This work was supported by JSPS KAKENHI Grant Number 19H01933, 20K14532, 21J00153, 21K03614, 21K03633, 21H04499, 22H01259 and Initiative on Promotion of Supercomputing for Young or Women Researchers, Information Technology centre, The University of Tokyo, and MEXT as “Program for Promoting Researches on the Supercomputer Fugaku” (Toward a unified view of the universe: from large scale structures to planets, Revealing the formation history of the universe with large-scale simulations and astronomical big data).
MF was supported by The University of Tokyo Excellent Young Researcher Program.
L.W. thanks the support from the one-hundred-talent project of Sun Yat-sen University
and the National Natural Science Foundation of China through grant 12073090.
L.W. also thanks the financial support from JSPS International Research Fellow (Graduate School of Science, The University of Tokyo).

\section*{Data Availability}
The data underlying this article will be shared on reasonable request to the corresponding author.
\textsc{petar} is available here: {\tt https://github.com/lwang-astro/PeTar}.
\textsc{celib} is available here: {\tt https://bitbucket.org/tsaitoh/celib}.



\bibliographystyle{mnras}
\bibliography{reference} 

\begin{thebibliography}{}
\makeatletter
\relax
\def\mn@urlcharsother{\let\do\@makeother \do\$\do\&\do\#\do\^\do\_\do\%\do\~}
\def\mn@doi{\begingroup\mn@urlcharsother \@ifnextchar [ {\mn@doi@}
  {\mn@doi@[]}}
\def\mn@doi@[#1]#2{\def\@tempa{#1}\ifx\@tempa\@empty \href
  {http://dx.doi.org/#2} {doi:#2}\else \href {http://dx.doi.org/#2} {#1}\fi
  \endgroup}
\def\mn@eprint#1#2{\mn@eprint@#1:#2::\@nil}
\def\mn@eprint@arXiv#1{\href {http://arxiv.org/abs/#1} {{\tt arXiv:#1}}}
\def\mn@eprint@dblp#1{\href {http://dblp.uni-trier.de/rec/bibtex/#1.xml}
  {dblp:#1}}
\def\mn@eprint@#1:#2:#3:#4\@nil{\def\@tempa {#1}\def\@tempb {#2}\def\@tempc
  {#3}\ifx \@tempc \@empty \let \@tempc \@tempb \let \@tempb \@tempa \fi \ifx
  \@tempb \@empty \def\@tempb {arXiv}\fi \@ifundefined
  {mn@eprint@\@tempb}{\@tempb:\@tempc}{\expandafter \expandafter \csname
  mn@eprint@\@tempb\endcsname \expandafter{\@tempc}}}

\bibitem[\protect\citeauthoryear{{Andr{\'e}} et~al.,}{{Andr{\'e}}
  et~al.}{2010}]{2010A&A...518L.102A}
{Andr{\'e}} P.,  et~al., 2010, \mn@doi [\aap] {10.1051/0004-6361/201014666},
  \href {http://adsabs.harvard.edu/abs/2010A%26A...518L.102A} {518, L102}

\bibitem[\protect\citeauthoryear{{Barnes} \& {Hut}}{{Barnes} \&
  {Hut}}{1986}]{Barnes1986}
{Barnes} J.,  {Hut} P.,  1986, \mn@doi [\nat] {10.1038/324446a0}, \href
  {https://ui.adsabs.harvard.edu/abs/1986Natur.324..446B} {324, 446}

\bibitem[\protect\citeauthoryear{{Bate}}{{Bate}}{2009}]{2009MNRAS.392..590B}
{Bate} M.~R.,  2009, \mn@doi [\mnras] {10.1111/j.1365-2966.2008.14106.x}, \href
  {https://ui.adsabs.harvard.edu/abs/2009MNRAS.392..590B} {392, 590}

\bibitem[\protect\citeauthoryear{{Beccari} et~al.,}{{Beccari}
  et~al.}{2017}]{2017A&A...604A..22B}
{Beccari} G.,  et~al., 2017, \mn@doi [\aap] {10.1051/0004-6361/201730432},
  \href {https://ui.adsabs.harvard.edu/abs/2017A&A...604A..22B} {604, A22}

\bibitem[\protect\citeauthoryear{{Bekki}}{{Bekki}}{2017}]{2017MNRAS.467.1857B}
{Bekki} K.,  2017, \mn@doi [\mnras] {10.1093/mnras/stx110}, \href
  {https://ui.adsabs.harvard.edu/abs/2017MNRAS.467.1857B} {467, 1857}

\bibitem[\protect\citeauthoryear{{Bonnell}, {Bate}  \& {Vine}}{{Bonnell}
  et~al.}{2003}]{2003MNRAS.343..413B}
{Bonnell} I.~A.,  {Bate} M.~R.,   {Vine} S.~G.,  2003, \mn@doi [\mnras]
  {10.1046/j.1365-8711.2003.06687.x}, \href
  {https://ui.adsabs.harvard.edu/abs/2003MNRAS.343..413B} {343, 413}

\bibitem[\protect\citeauthoryear{{Casertano} \& {Hut}}{{Casertano} \&
  {Hut}}{1985}]{1985ApJ...298...80C}
{Casertano} S.,  {Hut} P.,  1985, \mn@doi [\apj] {10.1086/163589}, \href
  {http://ads.nao.ac.jp/abs/1985ApJ...298...80C} {298, 80}

\bibitem[\protect\citeauthoryear{{Chen}, {Li}  \& {Vogelsberger}}{{Chen}
  et~al.}{2021}]{2021MNRAS.502.6157C}
{Chen} Y.,  {Li} H.,   {Vogelsberger} M.,  2021, \mn@doi [\mnras]
  {10.1093/mnras/stab491}, \href
  {https://ui.adsabs.harvard.edu/abs/2021MNRAS.502.6157C} {502, 6157}

\bibitem[\protect\citeauthoryear{{Da Rio}, {Robberto}, {Soderblom}, {Panagia},
  {Hillenbrand}, {Palla}  \& {Stassun}}{{Da Rio}
  et~al.}{2010}]{2010ApJ...722.1092D}
{Da Rio} N.,  {Robberto} M.,  {Soderblom} D.~R.,  {Panagia} N.,  {Hillenbrand}
  L.~A.,  {Palla} F.,   {Stassun} K.~G.,  2010, \mn@doi [\apj]
  {10.1088/0004-637X/722/2/1092}, \href
  {https://ui.adsabs.harvard.edu/abs/2010ApJ...722.1092D} {722, 1092}

\bibitem[\protect\citeauthoryear{{Da Rio}, {Tan}  \& {Jaehnig}}{{Da Rio}
  et~al.}{2014}]{2014ApJ...795...55D}
{Da Rio} N.,  {Tan} J.~C.,   {Jaehnig} K.,  2014, \mn@doi [\apj]
  {10.1088/0004-637X/795/1/55}, \href
  {https://ui.adsabs.harvard.edu/abs/2014ApJ...795...55D} {795, 55}

\bibitem[\protect\citeauthoryear{{Dekel} \& {Krumholz}}{{Dekel} \&
  {Krumholz}}{2013}]{2013MNRAS.432..455D}
{Dekel} A.,  {Krumholz} M.~R.,  2013, \mn@doi [\mnras] {10.1093/mnras/stt480},
  \href {https://ui.adsabs.harvard.edu/abs/2013MNRAS.432..455D} {432, 455}

\bibitem[\protect\citeauthoryear{{Eisenstein} \& {Hut}}{{Eisenstein} \&
  {Hut}}{1998}]{1998ApJ...498..137E}
{Eisenstein} D.~J.,  {Hut} P.,  1998, \mn@doi [\apj] {10.1086/305535}, \href
  {http://adsabs.harvard.edu/abs/1998ApJ...498..137E} {498, 137}

\bibitem[\protect\citeauthoryear{{Ferland}, {Korista}, {Verner}, {Ferguson},
  {Kingdon}  \& {Verner}}{{Ferland} et~al.}{1998}]{1998PASP..110..761F}
{Ferland} G.~J.,  {Korista} K.~T.,  {Verner} D.~A.,  {Ferguson} J.~W.,
  {Kingdon} J.~B.,   {Verner} E.~M.,  1998, \mn@doi [\pasp] {10.1086/316190},
  \href {http://adsabs.harvard.edu/abs/1998PASP..110..761F} {110, 761}

\bibitem[\protect\citeauthoryear{{Ferland} et~al.,}{{Ferland}
  et~al.}{2013}]{2013RMxAA..49..137F}
{Ferland} G.~J.,  et~al., 2013, Rev. Mex. Astron. Astrofis., \href
  {http://adsabs.harvard.edu/abs/2013RMxAA..49..137F} {49, 137}

\bibitem[\protect\citeauthoryear{{Ferland} et~al.,}{{Ferland}
  et~al.}{2017}]{2017RMxAA..53..385F}
{Ferland} G.~J.,  et~al., 2017, Rev. Mex. Astron. Astrofis., \href
  {https://ui.adsabs.harvard.edu/abs/2017RMxAA..53..385F} {53, 385}

\bibitem[\protect\citeauthoryear{{Fujii}}{{Fujii}}{2015}]{2015PASJ...67...59F}
{Fujii} M.~S.,  2015, \mn@doi [\pasj] {10.1093/pasj/psu137}, \href
  {https://ui.adsabs.harvard.edu/abs/2015PASJ...67...59F} {67, 59}

\bibitem[\protect\citeauthoryear{{Fujii}}{{Fujii}}{2019}]{2019MNRAS.486.3019F}
{Fujii} M.~S.,  2019, \mn@doi [\mnras] {10.1093/mnras/stz1056}, \href
  {https://ui.adsabs.harvard.edu/abs/2019MNRAS.486.3019F} {486, 3019}

\bibitem[\protect\citeauthoryear{{Fujii} \& {Portegies Zwart}}{{Fujii} \&
  {Portegies Zwart}}{2011}]{2011Sci...334.1380F}
{Fujii} M.~S.,  {Portegies Zwart} S.,  2011, \mn@doi [Science]
  {10.1126/science.1211927}, \href
  {http://adsabs.harvard.edu/abs/2011Sci...334.1380F} {334, 1380}

\bibitem[\protect\citeauthoryear{{Fujii}, {Iwasawa}, {Funato}  \&
  {Makino}}{{Fujii} et~al.}{2007}]{2007PASJ...59.1095F}
{Fujii} M.,  {Iwasawa} M.,  {Funato} Y.,   {Makino} J.,  2007, \pasj, \href
  {http://ads.nao.ac.jp/abs/2007PASJ...59.1095F} {59, 1095}

\bibitem[\protect\citeauthoryear{{Fujii}, {Saitoh}  \& {Portegies
  Zwart}}{{Fujii} et~al.}{2012}]{2012ApJ...753...85F}
{Fujii} M.~S.,  {Saitoh} T.~R.,   {Portegies Zwart} S.~F.,  2012, \mn@doi
  [\apj] {10.1088/0004-637X/753/1/85}, \href
  {https://ui.adsabs.harvard.edu/abs/2012ApJ...753...85F} {753, 85}

\bibitem[\protect\citeauthoryear{{Fujii}, {Saitoh}, {Wang}  \& {Hirai}}{{Fujii}
  et~al.}{2021a}]{2021PASJ...73.1057F}
{Fujii} M.~S.,  {Saitoh} T.~R.,  {Wang} L.,   {Hirai} Y.,  2021a, \mn@doi
  [\pasj] {10.1093/pasj/psab037}, \href
  {https://ui.adsabs.harvard.edu/abs/2021PASJ...73.1057F} {73, 1057}

\bibitem[\protect\citeauthoryear{{Fujii}, {Saitoh}, {Hirai}  \& {Wang}}{{Fujii}
  et~al.}{2021b}]{2021PASJ...73.1074F}
{Fujii} M.~S.,  {Saitoh} T.~R.,  {Hirai} Y.,   {Wang} L.,  2021b, \mn@doi
  [\pasj] {10.1093/pasj/psab061}, \href
  {https://ui.adsabs.harvard.edu/abs/2021PASJ...73.1074F} {73, 1074}

\bibitem[\protect\citeauthoryear{{Gaia Collaboration} et~al.,}{{Gaia
  Collaboration} et~al.}{2018}]{2018A&A...616A...1G}
{Gaia Collaboration} et~al., 2018, \mn@doi [\aap]
  {10.1051/0004-6361/201833051}, \href
  {https://ui.adsabs.harvard.edu/abs/2018A&A...616A...1G} {616, A1}

\bibitem[\protect\citeauthoryear{{Getman}, {Feigelson}  \& {Kuhn}}{{Getman}
  et~al.}{2014}]{2014ApJ...787..109G}
{Getman} K.~V.,  {Feigelson} E.~D.,   {Kuhn} M.~A.,  2014, \mn@doi [\apj]
  {10.1088/0004-637X/787/2/109}, \href
  {https://ui.adsabs.harvard.edu/abs/2014ApJ...787..109G} {787, 109}

\bibitem[\protect\citeauthoryear{{Gro{\ss}schedl} et~al.,}{{Gro{\ss}schedl}
  et~al.}{2018}]{2018A&A...619A.106G}
{Gro{\ss}schedl} J.~E.,  et~al., 2018, \mn@doi [\aap]
  {10.1051/0004-6361/201833901}, \href
  {https://ui.adsabs.harvard.edu/abs/2018A&A...619A.106G} {619, A106}

\bibitem[\protect\citeauthoryear{{Grudi{\'c}}, {Guszejnov}, {Hopkins},
  {Lamberts}, {Boylan-Kolchin}, {Murray}  \& {Schmitz}}{{Grudi{\'c}}
  et~al.}{2018}]{2018MNRAS.481..688G}
{Grudi{\'c}} M.~Y.,  {Guszejnov} D.,  {Hopkins} P.~F.,  {Lamberts} A.,
  {Boylan-Kolchin} M.,  {Murray} N.,   {Schmitz} D.,  2018, \mn@doi [\mnras]
  {10.1093/mnras/sty2303}, \href
  {https://ui.adsabs.harvard.edu/abs/2018MNRAS.481..688G} {481, 688}

\bibitem[\protect\citeauthoryear{{Grudi{\'c}}, {Kruijssen},
  {Faucher-Gigu{\`e}re}, {Hopkins}, {Ma}, {Quataert}  \&
  {Boylan-Kolchin}}{{Grudi{\'c}} et~al.}{2021}]{2021MNRAS.506.3239G}
{Grudi{\'c}} M.~Y.,  {Kruijssen} J.~M.~D.,  {Faucher-Gigu{\`e}re} C.-A.,
  {Hopkins} P.~F.,  {Ma} X.,  {Quataert} E.,   {Boylan-Kolchin} M.,  2021,
  \mn@doi [\mnras] {10.1093/mnras/stab1894}, \href
  {https://ui.adsabs.harvard.edu/abs/2021MNRAS.506.3239G} {506, 3239}

\bibitem[\protect\citeauthoryear{{Hillenbrand} \& {Hartmann}}{{Hillenbrand} \&
  {Hartmann}}{1998}]{1998ApJ...492..540H}
{Hillenbrand} L.~A.,  {Hartmann} L.~W.,  1998, \mn@doi [\apj] {10.1086/305076},
  \href {https://ui.adsabs.harvard.edu/abs/1998ApJ...492..540H} {492, 540}

\bibitem[\protect\citeauthoryear{{Hirai}, {Fujii}  \& {Saitoh}}{{Hirai}
  et~al.}{2021}]{2021PASJ...73.1036H}
{Hirai} Y.,  {Fujii} M.~S.,   {Saitoh} T.~R.,  2021, \mn@doi [\pasj]
  {10.1093/pasj/psab038}, \href
  {https://ui.adsabs.harvard.edu/abs/2021PASJ...73.1036H} {73, 1036}

\bibitem[\protect\citeauthoryear{{Howard}, {Pudritz}  \& {Harris}}{{Howard}
  et~al.}{2018}]{2018NatAs...2..725H}
{Howard} C.~S.,  {Pudritz} R.~E.,   {Harris} W.~E.,  2018, \mn@doi [Nature
  Astronomy] {10.1038/s41550-018-0506-0}, \href
  {https://ui.adsabs.harvard.edu/abs/2018NatAs...2..725H} {2, 725}

\bibitem[\protect\citeauthoryear{{Iwasawa}, {Portegies Zwart}  \&
  {Makino}}{{Iwasawa} et~al.}{2015}]{2015ComAC...2....6I}
{Iwasawa} M.,  {Portegies Zwart} S.,   {Makino} J.,  2015, \mn@doi
  [Computational Astrophysics and Cosmology] {10.1186/s40668-015-0010-1}, \href
  {https://ui.adsabs.harvard.edu/abs/2015ComAC...2....6I} {2, 6}

\bibitem[\protect\citeauthoryear{{Iwasawa}, {Tanikawa}, {Hosono}, {Nitadori},
  {Muranushi}  \& {Makino}}{{Iwasawa} et~al.}{2016}]{Iwasawa2016FDPS}
{Iwasawa} M.,  {Tanikawa} A.,  {Hosono} N.,  {Nitadori} K.,  {Muranushi} T.,
  {Makino} J.,  2016, \mn@doi [\pasj] {10.1093/pasj/psw053}, \href
  {https://ui.adsabs.harvard.edu/abs/2016PASJ...68...54I} {68, 54}

\bibitem[\protect\citeauthoryear{{Jerabkova}, {Beccari}, {Boffin},
  {Petr-Gotzens}, {Manara}, {Prada Moroni}, {Tognelli}  \&
  {Degl'Innocenti}}{{Jerabkova} et~al.}{2019}]{2019A&A...627A..57J}
{Jerabkova} T.,  {Beccari} G.,  {Boffin} H. M.~J.,  {Petr-Gotzens} M.~G.,
  {Manara} C.~F.,  {Prada Moroni} P.~G.,  {Tognelli} E.,   {Degl'Innocenti} S.,
   2019, \mn@doi [\aap] {10.1051/0004-6361/201935016}, \href
  {https://ui.adsabs.harvard.edu/abs/2019A&A...627A..57J} {627, A57}

\bibitem[\protect\citeauthoryear{{Jones} \& {Walker}}{{Jones} \&
  {Walker}}{1988}]{1988AJ.....95.1755J}
{Jones} B.~F.,  {Walker} M.~F.,  1988, \mn@doi [\aj] {10.1086/114773}, \href
  {https://ui.adsabs.harvard.edu/abs/1988AJ.....95.1755J} {95, 1755}

\bibitem[\protect\citeauthoryear{{Kim}, {Kim}  \& {Ostriker}}{{Kim}
  et~al.}{2016}]{2016ApJ...819..137K}
{Kim} J.-G.,  {Kim} W.-T.,   {Ostriker} E.~C.,  2016, \mn@doi [\apj]
  {10.3847/0004-637X/819/2/137}, \href
  {https://ui.adsabs.harvard.edu/abs/2016ApJ...819..137K} {819, 137}

\bibitem[\protect\citeauthoryear{{Kim}, {Lu}, {Konopacky}, {Chu}, {Toller},
  {Anderson}, {Theissen}  \& {Morris}}{{Kim}
  et~al.}{2019}]{2019AJ....157..109K}
{Kim} D.,  {Lu} J.~R.,  {Konopacky} Q.,  {Chu} L.,  {Toller} E.,  {Anderson}
  J.,  {Theissen} C.~A.,   {Morris} M.~R.,  2019, \mn@doi [\aj]
  {10.3847/1538-3881/aafb09}, \href
  {https://ui.adsabs.harvard.edu/abs/2019AJ....157..109K} {157, 109}

\bibitem[\protect\citeauthoryear{{Kinoshita}, {Yoshida}  \&
  {Nakai}}{{Kinoshita} et~al.}{1991}]{1991CeMDA..50...59K}
{Kinoshita} H.,  {Yoshida} H.,   {Nakai} H.,  1991, Celestial Mechanics and
  Dynamical Astronomy, \href {http://ads.nao.ac.jp/abs/1991CeMDA..50...59K}
  {50, 59}

\bibitem[\protect\citeauthoryear{{Kounkel} et~al.,}{{Kounkel}
  et~al.}{2017}]{2017ApJ...834..142K}
{Kounkel} M.,  et~al., 2017, \mn@doi [\apj] {10.3847/1538-4357/834/2/142},
  \href {https://ui.adsabs.harvard.edu/abs/2017ApJ...834..142K} {834, 142}

\bibitem[\protect\citeauthoryear{{Kroupa}}{{Kroupa}}{2001}]{2001MNRAS.322..231K}
{Kroupa} P.,  2001, \mn@doi [\mnras] {10.1046/j.1365-8711.2001.04022.x}, \href
  {http://adsabs.harvard.edu/abs/2001MNRAS.322..231K} {322, 231}

\bibitem[\protect\citeauthoryear{{Kroupa}, {Je{\v{r}}{\'a}bkov{\'a}},
  {Dinnbier}, {Beccari}  \& {Yan}}{{Kroupa} et~al.}{2018}]{2018A&A...612A..74K}
{Kroupa} P.,  {Je{\v{r}}{\'a}bkov{\'a}} T.,  {Dinnbier} F.,  {Beccari} G.,
  {Yan} Z.,  2018, \mn@doi [\aap] {10.1051/0004-6361/201732151}, \href
  {https://ui.adsabs.harvard.edu/abs/2018A&A...612A..74K} {612, A74}

\bibitem[\protect\citeauthoryear{{Krumholz} \& {Tan}}{{Krumholz} \&
  {Tan}}{2007}]{2007ApJ...654..304K}
{Krumholz} M.~R.,  {Tan} J.~C.,  2007, \mn@doi [\apj] {10.1086/509101}, \href
  {https://ui.adsabs.harvard.edu/abs/2007ApJ...654..304K} {654, 304}

\bibitem[\protect\citeauthoryear{{Krumholz}, {Klein}  \& {McKee}}{{Krumholz}
  et~al.}{2012}]{2012ApJ...754...71K}
{Krumholz} M.~R.,  {Klein} R.~I.,   {McKee} C.~F.,  2012, \mn@doi [\apj]
  {10.1088/0004-637X/754/1/71}, \href
  {http://adsabs.harvard.edu/abs/2012ApJ...754...71K} {754, 71}

\bibitem[\protect\citeauthoryear{{Kuhn}, {Hillenbrand}, {Sills}, {Feigelson}
  \& {Getman}}{{Kuhn} et~al.}{2019}]{2019ApJ...870...32K}
{Kuhn} M.~A.,  {Hillenbrand} L.~A.,  {Sills} A.,  {Feigelson} E.~D.,   {Getman}
  K.~V.,  2019, \mn@doi [\apj] {10.3847/1538-4357/aaef8c}, \href
  {https://ui.adsabs.harvard.edu/abs/2019ApJ...870...32K} {870, 32}

\bibitem[\protect\citeauthoryear{{Lah{\'e}n}, {Naab}, {Johansson}, {Elmegreen},
  {Hu}, {Walch}, {Steinwand el}  \& {Moster}}{{Lah{\'e}n}
  et~al.}{2020}]{2020ApJ...891....2L}
{Lah{\'e}n} N.,  {Naab} T.,  {Johansson} P.~H.,  {Elmegreen} B.,  {Hu} C.-Y.,
  {Walch} S.,  {Steinwand el} U.~P.,   {Moster} B.~P.,  2020, \mn@doi [\apj]
  {10.3847/1538-4357/ab7190}, \href
  {https://ui.adsabs.harvard.edu/abs/2020ApJ...891....2L} {891, 2}

\bibitem[\protect\citeauthoryear{{Lanz} \& {Hubeny}}{{Lanz} \&
  {Hubeny}}{2003}]{2003ApJS..146..417L}
{Lanz} T.,  {Hubeny} I.,  2003, \mn@doi [\apjs] {10.1086/374373}, \href
  {https://ui.adsabs.harvard.edu/abs/2003ApJS..146..417L} {146, 417}

\bibitem[\protect\citeauthoryear{{McBride} \& {Kounkel}}{{McBride} \&
  {Kounkel}}{2019}]{2019ApJ...884....6M}
{McBride} A.,  {Kounkel} M.,  2019, \mn@doi [\apj] {10.3847/1538-4357/ab3df9},
  \href {https://ui.adsabs.harvard.edu/abs/2019ApJ...884....6M} {884, 6}

\bibitem[\protect\citeauthoryear{{Menten}, {Reid}, {Forbrich}  \&
  {Brunthaler}}{{Menten} et~al.}{2007}]{2007A&A...474..515M}
{Menten} K.~M.,  {Reid} M.~J.,  {Forbrich} J.,   {Brunthaler} A.,  2007,
  \mn@doi [\aap] {10.1051/0004-6361:20078247}, \href
  {https://ui.adsabs.harvard.edu/abs/2007A&A...474..515M} {474, 515}

\bibitem[\protect\citeauthoryear{{Nitadori} \& {Makino}}{{Nitadori} \&
  {Makino}}{2008}]{2008NewA...13..498N}
{Nitadori} K.,  {Makino} J.,  2008, \mn@doi [New Astronomy]
  {10.1016/j.newast.2008.01.010}, \href
  {http://ads.nao.ac.jp/abs/2008NewA...13..498N} {13, 498}

\bibitem[\protect\citeauthoryear{{Oshino}, {Funato}  \& {Makino}}{{Oshino}
  et~al.}{2011}]{2011PASJ...63..881O}
{Oshino} S.,  {Funato} Y.,   {Makino} J.,  2011, \mn@doi [\pasj]
  {10.1093/pasj/63.4.881}, \href
  {https://ui.adsabs.harvard.edu/abs/2011PASJ...63..881O} {63, 881}

\bibitem[\protect\citeauthoryear{{Palla} \& {Stahler}}{{Palla} \&
  {Stahler}}{2000}]{2000ApJ...540..255P}
{Palla} F.,  {Stahler} S.~W.,  2000, \mn@doi [\apj] {10.1086/309312}, \href
  {https://ui.adsabs.harvard.edu/abs/2000ApJ...540..255P} {540, 255}

\bibitem[\protect\citeauthoryear{{Pelupessy}, {van Elteren}, {de Vries},
  {McMillan}, {Drost}  \& {Portegies Zwart}}{{Pelupessy}
  et~al.}{2013}]{2013A&A...557A..84P}
{Pelupessy} F.~I.,  {van Elteren} A.,  {de Vries} N.,  {McMillan} S.~L.~W.,
  {Drost} N.,   {Portegies Zwart} S.~F.,  2013, \mn@doi [\aap]
  {10.1051/0004-6361/201321252}, \href
  {http://adsabs.harvard.edu/abs/2013A%26A...557A..84P} {557, A84}

\bibitem[\protect\citeauthoryear{{Platais} et~al.,}{{Platais}
  et~al.}{2020}]{2020AJ....159..272P}
{Platais} I.,  et~al., 2020, \mn@doi [\aj] {10.3847/1538-3881/ab8d42}, \href
  {https://ui.adsabs.harvard.edu/abs/2020AJ....159..272P} {159, 272}

\bibitem[\protect\citeauthoryear{{Portegies Zwart} \& {McMillan}}{{Portegies
  Zwart} \& {McMillan}}{2018}]{AMUSE}
{Portegies Zwart} S.,  {McMillan} S.,  2018, {Astrophysical Recipes: the Art of
  AMUSE}.
AAS IOP Astronomy

\bibitem[\protect\citeauthoryear{{Portegies Zwart}, {McMillan}, {van Elteren},
  {Pelupessy}  \& {de Vries}}{{Portegies Zwart}
  et~al.}{2013}]{2013CoPhC.183..456P}
{Portegies Zwart} S.,  {McMillan} S.~L.~W.,  {van Elteren} E.,  {Pelupessy} I.,
    {de Vries} N.,  2013, \mn@doi [Computer Physics Communications]
  {10.1016/j.cpc.2012.09.024}, \href
  {http://adsabs.harvard.edu/abs/2013CoPhC.183..456P} {183, 456}

\bibitem[\protect\citeauthoryear{{Renaud} et~al.,}{{Renaud}
  et~al.}{2013}]{2013MNRAS.436.1836R}
{Renaud} F.,  et~al., 2013, \mn@doi [\mnras] {10.1093/mnras/stt1698}, \href
  {https://ui.adsabs.harvard.edu/abs/2013MNRAS.436.1836R} {436, 1836}

\bibitem[\protect\citeauthoryear{{Saitoh}}{{Saitoh}}{2017}]{2017AJ....153...85S}
{Saitoh} T.~R.,  2017, \mn@doi [\aj] {10.3847/1538-3881/153/2/85}, \href
  {http://adsabs.harvard.edu/abs/2017AJ....153...85S} {153, 85}

\bibitem[\protect\citeauthoryear{{Saitoh}, {Daisaka}, {Kokubo}, {Makino},
  {Okamoto}, {Tomisaka}, {Wada}  \& {Yoshida}}{{Saitoh}
  et~al.}{2008}]{2008PASJ...60..667S}
{Saitoh} T.~R.,  {Daisaka} H.,  {Kokubo} E.,  {Makino} J.,  {Okamoto} T.,
  {Tomisaka} K.,  {Wada} K.,   {Yoshida} N.,  2008, \mn@doi [\pasj]
  {10.1093/pasj/60.4.667}, \href
  {http://adsabs.harvard.edu/abs/2008PASJ...60..667S} {60, 667}

\bibitem[\protect\citeauthoryear{{Saitoh}, {Daisaka}, {Kokubo}, {Makino},
  {Okamoto}, {Tomisaka}, {Wada}  \& {Yoshida}}{{Saitoh}
  et~al.}{2009}]{2009PASJ...61..481S}
{Saitoh} T.~R.,  {Daisaka} H.,  {Kokubo} E.,  {Makino} J.,  {Okamoto} T.,
  {Tomisaka} K.,  {Wada} K.,   {Yoshida} N.,  2009, \mn@doi [\pasj]
  {10.1093/pasj/61.3.481}, \href
  {http://adsabs.harvard.edu/abs/2009PASJ...61..481S} {61, 481}

\bibitem[\protect\citeauthoryear{{Sana} et~al.,}{{Sana}
  et~al.}{2012}]{2012Sci...337..444S}
{Sana} H.,  et~al., 2012, \mn@doi [Science] {10.1126/science.1223344}, \href
  {https://ui.adsabs.harvard.edu/abs/2012Sci...337..444S} {337, 444}

\bibitem[\protect\citeauthoryear{{Schoettler}, {de Bruijne}, {Vaher}  \&
  {Parker}}{{Schoettler} et~al.}{2020}]{2020MNRAS.495.3104S}
{Schoettler} C.,  {de Bruijne} J.,  {Vaher} E.,   {Parker} R.~J.,  2020,
  \mn@doi [\mnras] {10.1093/mnras/staa1228}, \href
  {https://ui.adsabs.harvard.edu/abs/2020MNRAS.495.3104S} {495, 3104}

\bibitem[\protect\citeauthoryear{{Str{\"o}mgren}}{{Str{\"o}mgren}}{1939}]{1939ApJ....89..526S}
{Str{\"o}mgren} B.,  1939, \mn@doi [\apj] {10.1086/144074}, \href
  {https://ui.adsabs.harvard.edu/abs/1939ApJ....89..526S} {89, 526}

\bibitem[\protect\citeauthoryear{{Tan}, {Krumholz}  \& {McKee}}{{Tan}
  et~al.}{2006}]{2006ApJ...641L.121T}
{Tan} J.~C.,  {Krumholz} M.~R.,   {McKee} C.~F.,  2006, \mn@doi [\apjl]
  {10.1086/504150}, \href
  {https://ui.adsabs.harvard.edu/abs/2006ApJ...641L.121T} {641, L121}

\bibitem[\protect\citeauthoryear{{Theissen}, {Konopacky}, {Lu}, {Kim}, {Zhang},
  {Hsu}, {Chu}  \& {Wei}}{{Theissen} et~al.}{2022}]{2022ApJ...926..141T}
{Theissen} C.~A.,  {Konopacky} Q.~M.,  {Lu} J.~R.,  {Kim} D.,  {Zhang} S.~Y.,
  {Hsu} C.-C.,  {Chu} L.,   {Wei} L.,  2022, \mn@doi [\apj]
  {10.3847/1538-4357/ac3252}, \href
  {https://ui.adsabs.harvard.edu/abs/2022ApJ...926..141T} {926, 141}

\bibitem[\protect\citeauthoryear{{V{\'a}zquez-Semadeni},
  {Gonz{\'a}lez-Samaniego}  \& {Col{\'\i}n}}{{V{\'a}zquez-Semadeni}
  et~al.}{2017}]{2017MNRAS.467.1313V}
{V{\'a}zquez-Semadeni} E.,  {Gonz{\'a}lez-Samaniego} A.,   {Col{\'\i}n} P.,
  2017, \mn@doi [\mnras] {10.1093/mnras/stw3229}, \href
  {https://ui.adsabs.harvard.edu/abs/2017MNRAS.467.1313V} {467, 1313}

\bibitem[\protect\citeauthoryear{{Wang}, {Kroupa}  \& {Jerabkova}}{{Wang}
  et~al.}{2019}]{2019MNRAS.484.1843W}
{Wang} L.,  {Kroupa} P.,   {Jerabkova} T.,  2019, \mn@doi [\mnras]
  {10.1093/mnras/sty2232}, \href
  {https://ui.adsabs.harvard.edu/abs/2019MNRAS.484.1843W} {484, 1843}

\bibitem[\protect\citeauthoryear{{Wang}, {Nitadori}  \& {Makino}}{{Wang}
  et~al.}{2020a}]{2020MNRAS.493.3398W}
{Wang} L.,  {Nitadori} K.,   {Makino} J.,  2020a, \mn@doi [\mnras]
  {10.1093/mnras/staa480}, \href
  {https://ui.adsabs.harvard.edu/abs/2020MNRAS.493.3398W} {493, 3398}

\bibitem[\protect\citeauthoryear{{Wang}, {Iwasawa}, {Nitadori}  \&
  {Makino}}{{Wang} et~al.}{2020b}]{2020MNRAS.497..536W}
{Wang} L.,  {Iwasawa} M.,  {Nitadori} K.,   {Makino} J.,  2020b, \mn@doi
  [\mnras] {10.1093/mnras/staa1915}, \href
  {https://ui.adsabs.harvard.edu/abs/2020MNRAS.497..536W} {497, 536}

\bibitem[\protect\citeauthoryear{{Wisdom} \& {Holman}}{{Wisdom} \&
  {Holman}}{1991}]{1991AJ....102.1528W}
{Wisdom} J.,  {Holman} M.,  1991, \mn@doi [\aj] {10.1086/115978}, \href
  {http://ads.nao.ac.jp/abs/1991AJ....102.1528W} {102, 1528}

\makeatother
\end{thebibliography}


\appendix
\section{Run-to-run variation}
It is known that the formation histories of star clusters are affected by the randomness of the initial turbulent velocity field. 
In this section, we present the results of additional simulations to show that clump mergers generally enhances the star formation and causes the dynamical evolution of star clusters. 
In Table~\ref{tb:IC2}, we summarize the parameters for the initial condition of the additional simulations. We adopted the mass resolution for the SPH particle of $0.1\,M_{\odot}$, which is an order of magnitude lower mass resolution than that of Model ONC. We call this model Model Low. The initial density of this model is higher than Model ONC. This results in earlier star formation. These settings can reduce the simulation costs. We performed three runs for this model changing the random seeds for the initial turbulent velocity field. We call them Runs s1, s2, and s3. 

\begin{figure}
	\includegraphics[width=0.95\columnwidth]{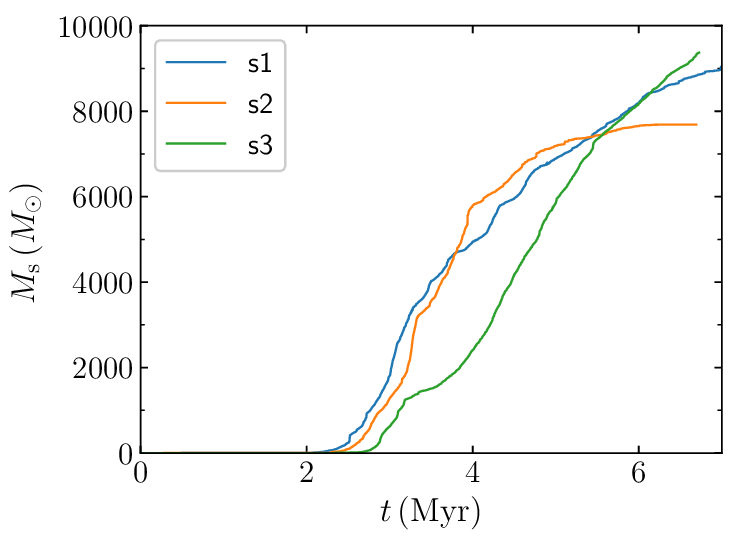}
	\caption{Stellar mass evolution for Models Low.}
    \label{fig:mass_evolution_B}
\end{figure}

In Figure~\ref{fig:mass_evolution_B}, we present the total stellar mass evolution of Model Low. Compared to Model ONC, more stars were formed in this model. The star formation ended at $t\sim6$--7\,Myr. We investigated the merger histories of these runs. We note that Run s1 has a relatively quiet formation history. There was only one massive merging clump with a mass of more than 200\,$M_{\odot}$. The other merging clumps were less than 100\,$M_{\odot}$.

\begin{figure*}
	\includegraphics[width=0.66\columnwidth]{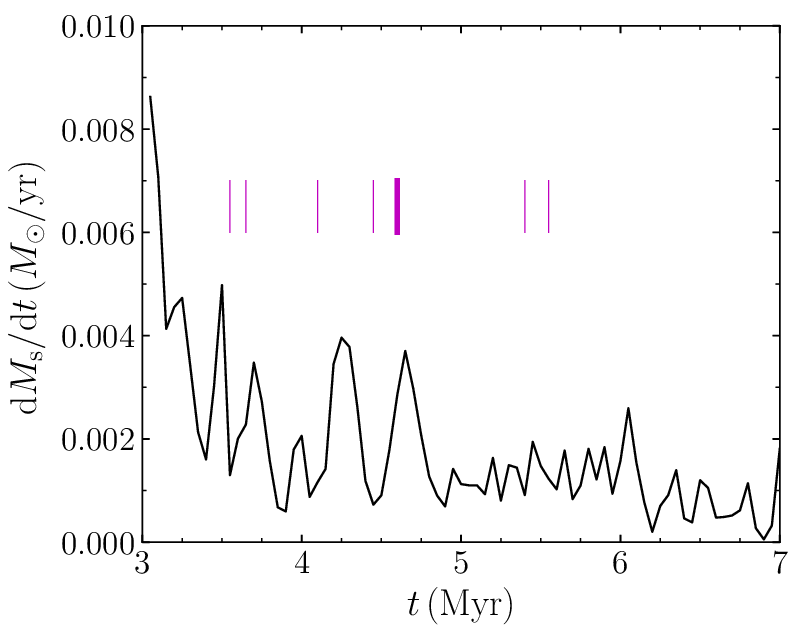}
	\includegraphics[width=0.66\columnwidth]{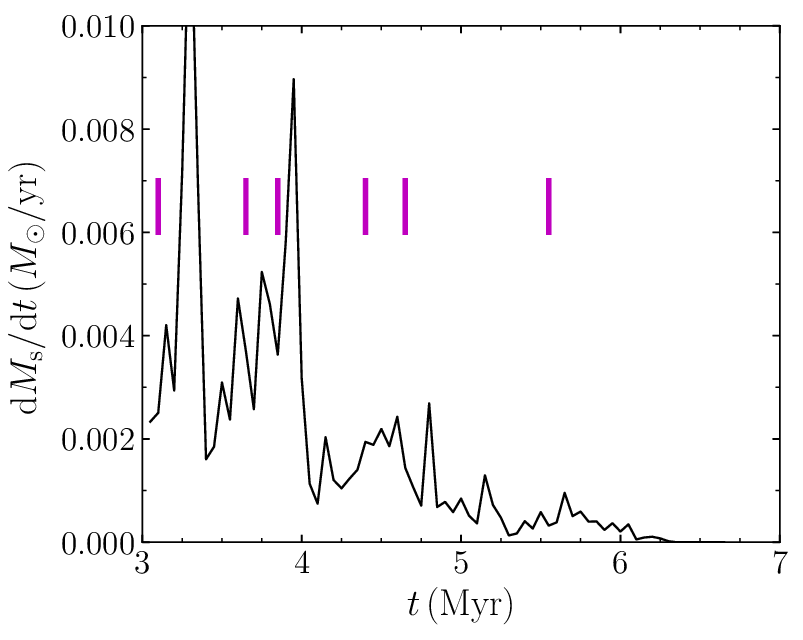}
	\includegraphics[width=0.66\columnwidth]{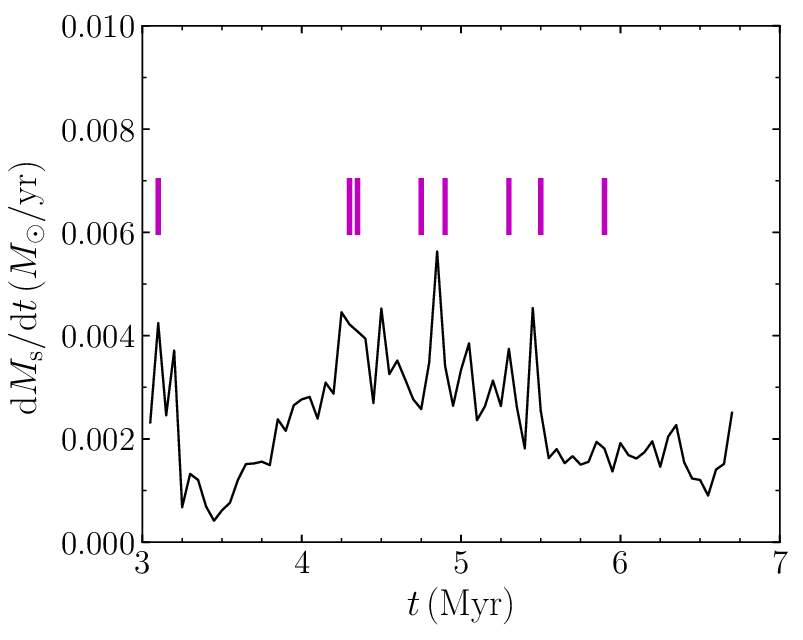}\\
	\includegraphics[width=0.66\columnwidth]{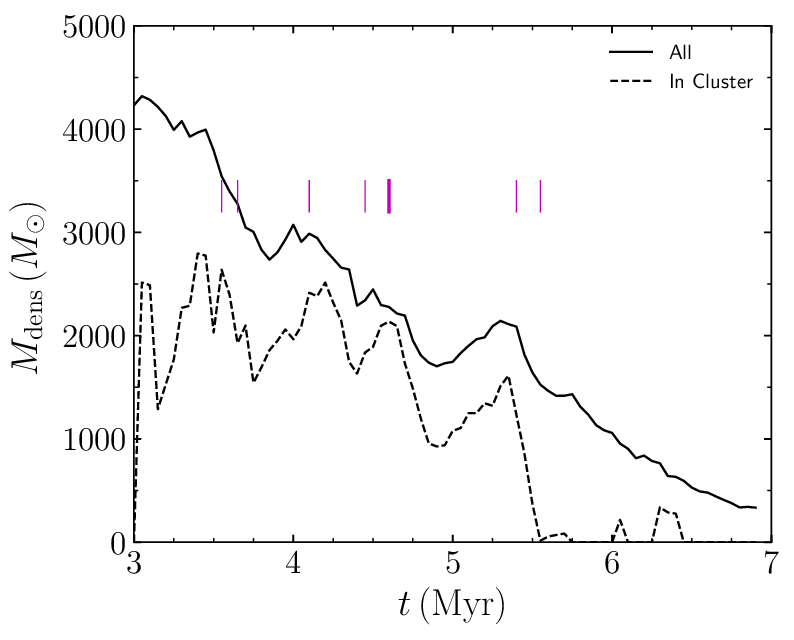}
	\includegraphics[width=0.66\columnwidth]{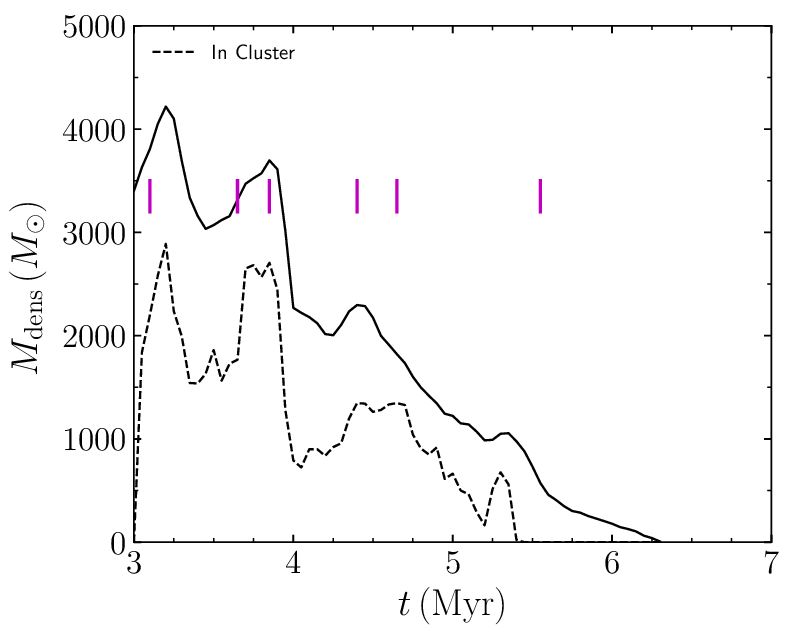}
	\includegraphics[width=0.66\columnwidth]{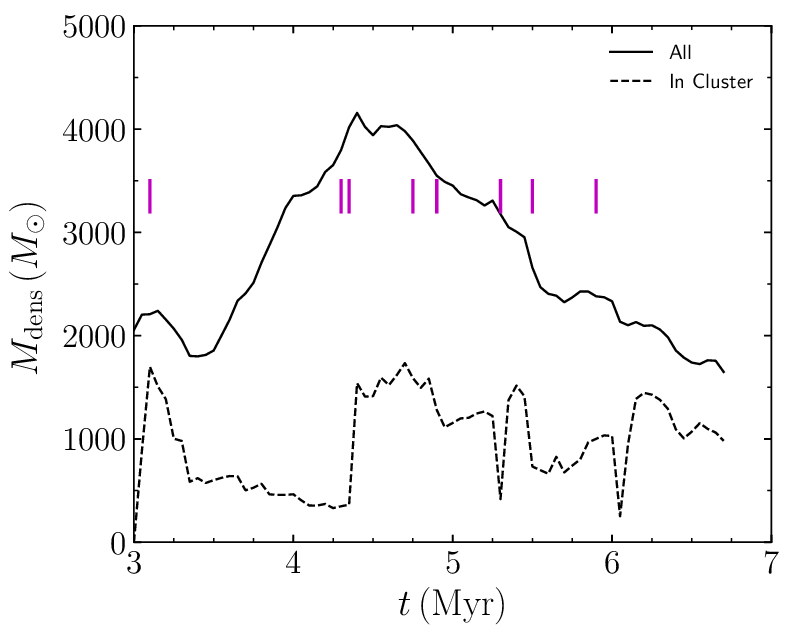}
	\caption{Star formation rate (top) and dense gas mass (bottom) for Runs S1, S2, and S3 (from left to right) of Model Low. Thick vertical lines indicate the timings of clump mergers with $>200M_{\odot}$ and thin lines indicate mergers with $100$--$200 M_{\odot}$.}
    \label{fig:dens_gas_mass_B}
\end{figure*}

The formation histories of these three runs have a large variation. In Figure~\ref{fig:dens_gas_mass_B}, we present the star formation rate and dense gas mass of these three runs with the merger times. As is shown in Figures~\ref{fig:star_formation_rate} and \ref{fig:dens_gas_mass}, the star formation increased after clump mergers due to the increase of dense gas in the cluster center. 

\begin{figure*}
	\includegraphics[width=0.66\columnwidth]{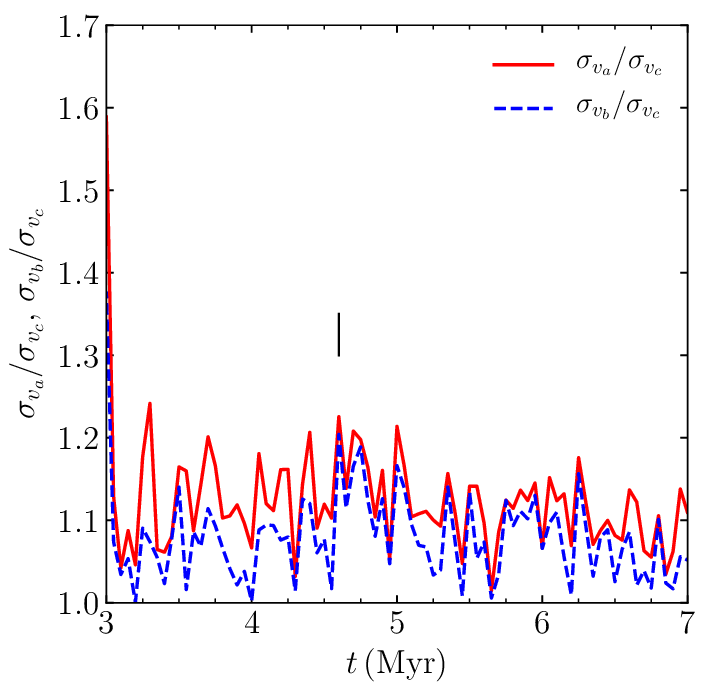}
	\includegraphics[width=0.66\columnwidth]{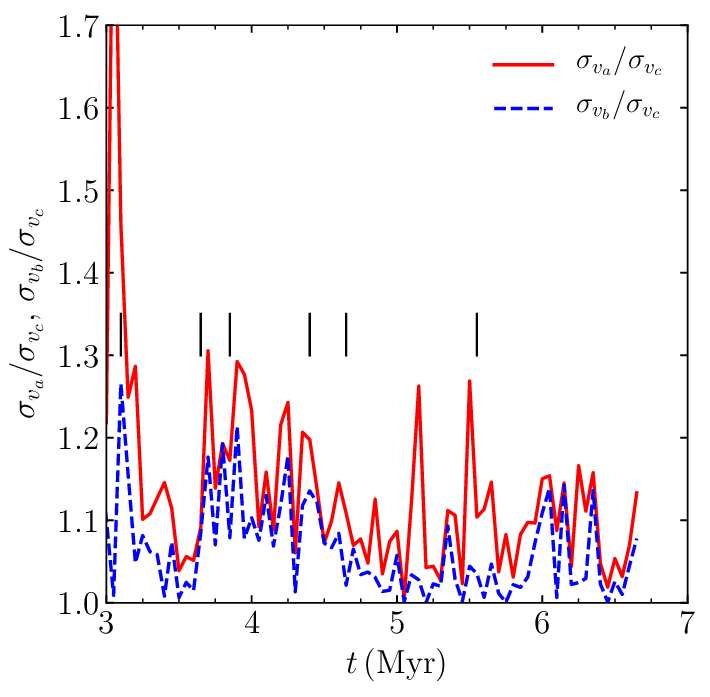}
	\includegraphics[width=0.66\columnwidth]{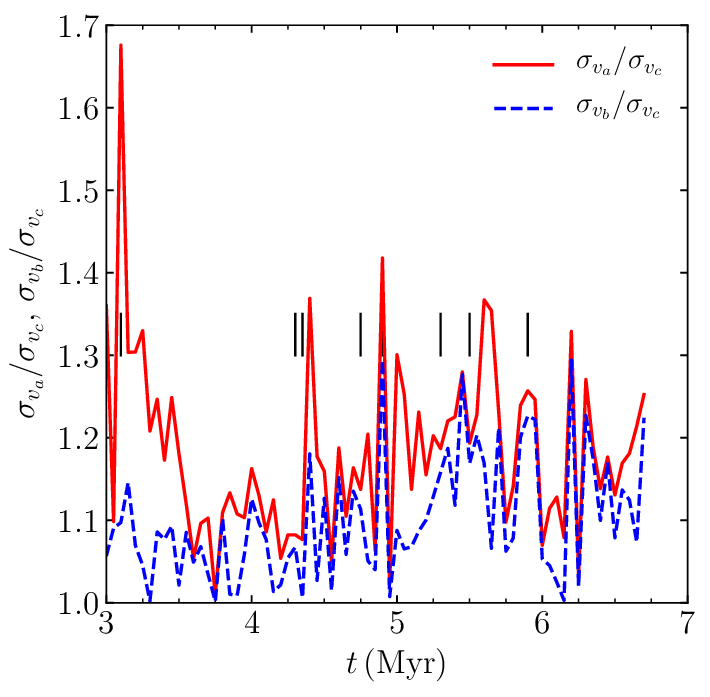}
	\caption{Velocity anisotropy for Runs s1, s2, and s3 (from left to right) of Model Low. Vertical lines indicate the timings of clump mergers with $>200M_{\odot}$.}
    \label{fig:vel_aniso_B}
\end{figure*}

\begin{figure*}
	\includegraphics[width=0.66\columnwidth]{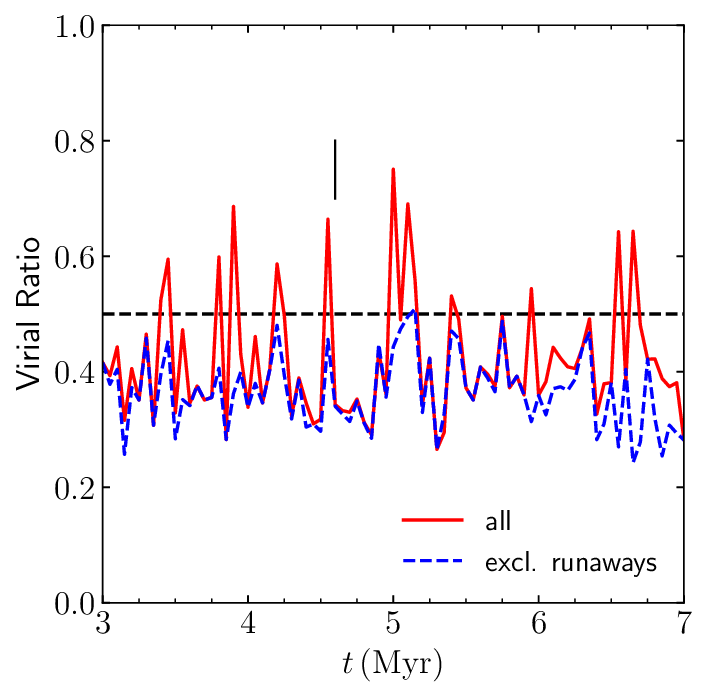}
	\includegraphics[width=0.66\columnwidth]{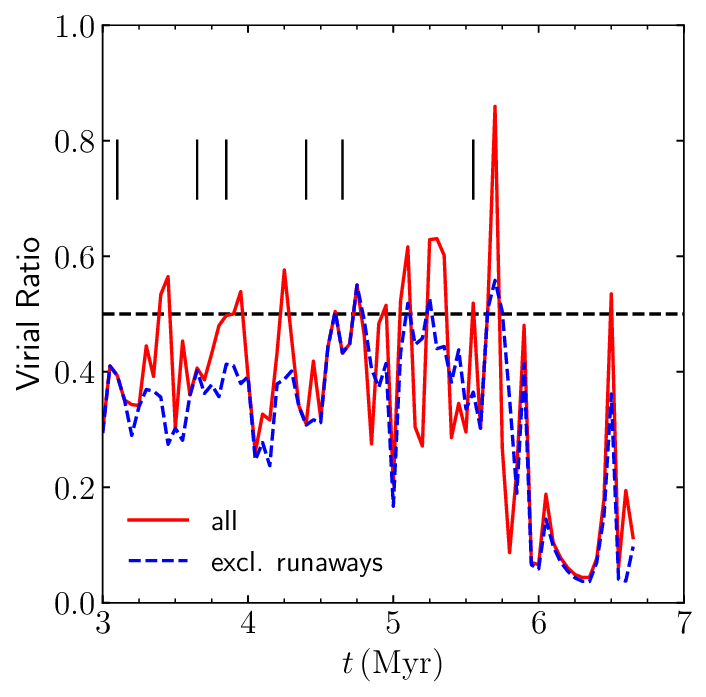}
	\includegraphics[width=0.66\columnwidth]{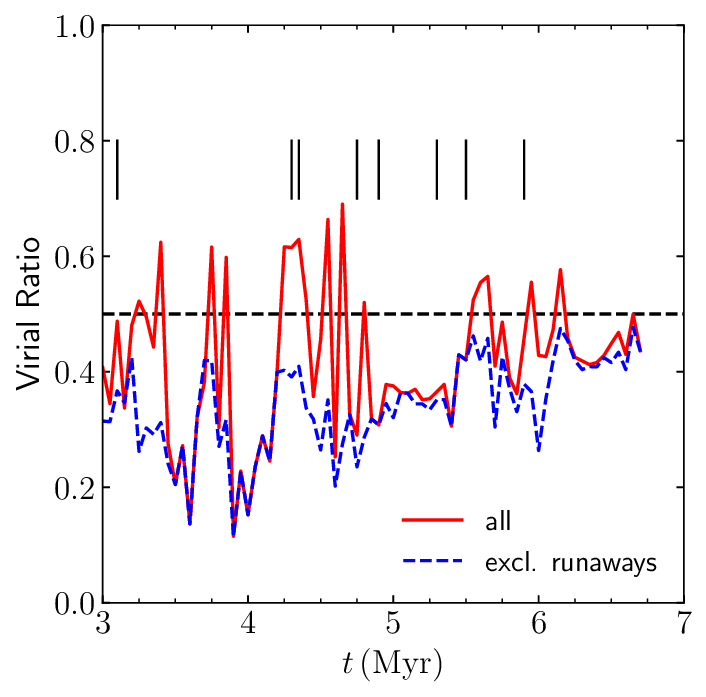}
	\caption{Virial ratio for Runs s1, s2, and s3 (from left to right) of Model Low. Vertical lines indicate the timings of clump mergers with $>200M_{\odot}$. Horizontal dashed line indicates 'virialized.'}
    \label{fig:virial_ratio_B}
\end{figure*}

We also present the time evolution of the velocity anisotropy and virial ratio for these runs in Figures~\ref{fig:vel_aniso_B} and \ref{fig:virial_ratio_B}, respectively. Similar to Model ONC (see Figures~\ref{fig:vel_aniso} and \ref{fig:virial_ratio}), the velocity anisotropy and virial ratio tend to increase after clump mergers. For Run s1, the velocity anisotropy was lower than the other runs. As mentioned above, Run s1 has experienced only one merger of relatively massive clump ($>200\,M_{\odot}$). Our result suggests that mergers of small clumps ($\lesssim100\,M_{\odot}$) is not efficient to cause a velocity anisotropy for several thousands-$M_{\odot}$ clusters.

\begin{figure*}
	\includegraphics[width=0.66\columnwidth]{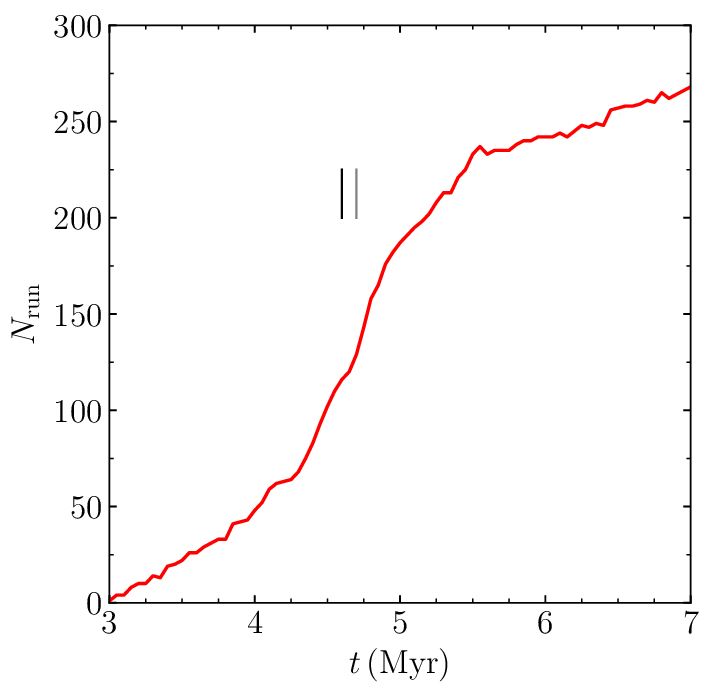}
	\includegraphics[width=0.66\columnwidth]{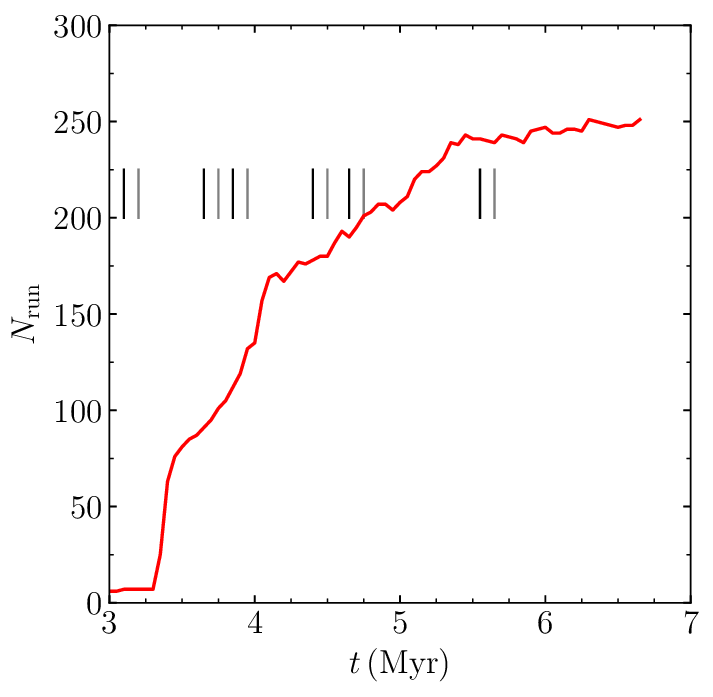}
	\includegraphics[width=0.66\columnwidth]{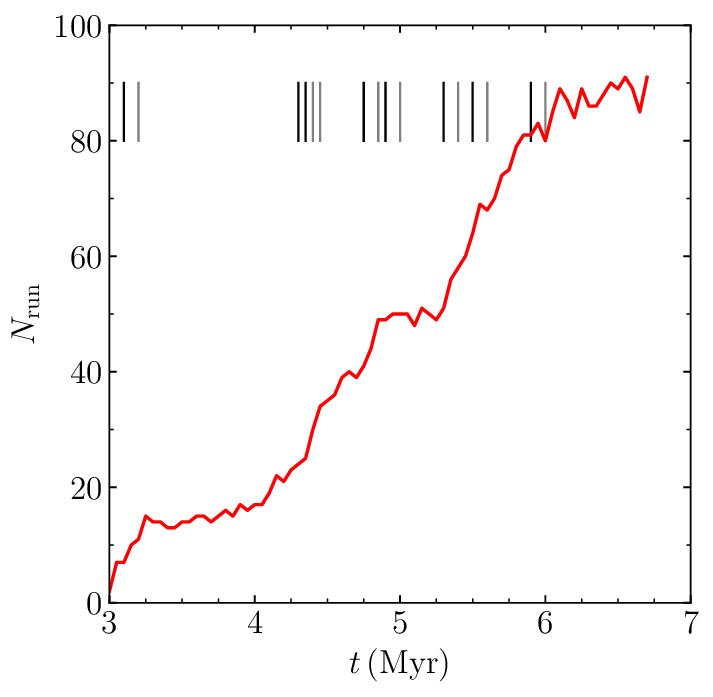}
	\caption{Same as Figure~\ref{fig:N_runaways} but for Runs S1, S2, and S3 (from left to right) of Model Low.}
    \label{fig:N_runaways_B}
\end{figure*}

In Figure~\ref{fig:N_runaways_B}, we present the cumulative number of runaway stars for Model Low. The rapid increases of runaway stars roughly match to the merger events. The formation histories of runaways have a run-to-run variation depending on the random seed for the initial turbulent velocity field. 

Thus, the formation histories of star clusters, such as merger histories, star formation histories, and resulting dynamical evolution, have a large variation due to the randomness of the turbulent velocity field even if we start the simulations using the same mass and density for the initial molecular cloud. However, the dynamical evolution of star clusters due to clump mergers is commonly observed.  

\begin{table*}
\begin{center}
\caption{Parameters for low-resolution models\label{tb:IC2}}
\begin{tabular}{lccccccccccccc}
\hline
   Name  & $M_{\rm g}$  & $m_{\rm g}$ & $R_{\rm g}$& $n_{\rm ini}$ & $t_{\rm ff,ini}$  & $\alpha_{\rm vir}$ & $\epsilon_{\rm g}$ & $\epsilon_{\rm s}$ & $n_{\rm th}$ & $r_{\rm max}$ & $\Delta t_{\rm B}$ & $\Delta t_{\rm soft}$ & $r_{\rm out}$ \\
       & $(M_{\odot})$ & $(M_{\odot})$ & (pc) &  (cm$^{-3}$) & (Myr) &  & (pc) & (pc) & (cm$^{-3}$) & (pc) & (yr) & & (pc)\\ 
      \hline
  Low & $2\times 10^4$ & $0.1$ & 7.84 & $287$ & $2.57$ & $0.5$  & $0.07$ & $0.0$ & $1.0\times 10^5$ & $0.2$ & $200$ & $\Delta t_{\rm B}/1024$ & 0.001 \\
\hline
\end{tabular}

From the left: model name, initial cloud mass ($M_{\rm g}$), gas-particle mass ($m_{\rm g}$), initial cloud radius ($R_{\rm g}$), initial cloud density ($n_{\rm ini}$), initial free-fall time ($t_{\rm ff, ini}$), initial virial ratio ($\alpha_{\rm vir}=|E_{\rm k}|/|E_{\rm p}|$), softening length for gas ($\epsilon_{\rm g}$) and stars ($\epsilon_{\rm s}$), star formation threshold density ($n_{\rm th}$), the maximum search radius for the star formation ($r_{\rm max}$), timestep for Bridge ($\Delta t_{\rm B}$) and soft part ($\Delta t_{\rm soft}$), and the outer cut-off radius for Bridge ($r_{\rm out}$).
\end{center}
\end{table*}

\bsp	
\label{lastpage}
\end{document}